\documentclass[apj,onecolumn,numberedappendix]{emulateapj}
\usepackage{graphicx}
\usepackage{dcolumn}
\usepackage{bm}
\usepackage{amssymb,amsmath}
\usepackage{latexsym}
\usepackage{color}
\usepackage[normalem]{ulem} 
\usepackage{cancel}

\allowdisplaybreaks

\begin{document}

\title{The stability of tidally deformed neutron stars to three- and four-mode coupling}

\author{Tejaswi Venumadhav, Aaron Zimmerman, and Christopher M. Hirata}
\affiliation{Theoretical Astrophysics, Mail Code 350-17, California Institute of Technology, Pasadena, California 91125, USA}
\date{\today}
\submitted{}

\begin{abstract}
It has recently been suggested that the tidal deformation of a neutron star excites daughter $p$- and $g$-modes to large amplitudes via a quasi-static instability. This would remove energy from the tidal bulge, resulting in dissipation and possibly affecting the phase evolution of inspiralling binary neutron stars and hence the extraction of binary parameters from gravitational wave observations.  This instability appears to arise because of a large three-mode interaction among the tidal mode and high-order $p$- and $g$-modes of similar radial wavenumber. We show that additional four-mode interactions enter into the analysis at the same order as the three-mode terms previously considered. We compute these four-mode couplings by finding a volume-preserving coordinate transformation that relates the energy of a tidally deformed star to that of a radially perturbed spherical star. Using this method, we relate the four-mode coupling to three-mode couplings and show that there is a near-exact cancellation between the destabilizing effect of the three-mode interactions and the stabilizing effect of the four-mode interaction. We then show that the equilibrium tide is stable against the quasi-static decay into daughter $p$- and $g$-modes to leading order. The leading deviation from the quasi-static approximation due to orbital motion of the binary is considered; while it may slightly spoil the near-cancellation, any resulting instability timescale is at least of order the gravitational-wave inspiral time. We conclude that the $p$-/$g$-mode coupling does not lead to a quasi-static instability, and does not impact the phase evolution of gravitational waves from binary neutron stars.
\end{abstract}

\keywords{stars: neutron --- stars: oscillations --- binaries: close}

\maketitle
 
\section{Introduction}
\label{sec:Intro}

Compact binary systems are thought to host some of the most energetic phenomena in the universe. In particular, neutron star binaries can be exceptionally bright sources of both gravitational and electromagnetic radiation during their inspiral and eventual merger. Such systems provide a unique window into a variety of fundamental physical processes. For example, merging binaries which host at least one neutron star are thought to be sources of short gamma ray bursts. These binary inspirals are also the most promising sources for the upcoming generation of gravitational wave detectors, such as Advanced LIGO \citep{aLIGO2010}, Advanced VIRGO \citep{aVIRGO2012}, and KAGRA \citep{KAGRA2012}. Observations of compact binaries through their gravitational wave emission should provide precise measurements of the binary parameters~\citep[e.g.][]{Cutler1994,Arun2005,LIGOParam2013}, including possibly the indirect measurement of the neutron star equation of state through the effects of tidal deformation of the binary companions on the gravitational waveform~\citep[e.g.][]{FlanaganHinderer2008,HindererLackey2010,Hotokezaka2013,Read2013} or the final cutoff frequency of the gravitational waveform~\citep[e.g.][]{Cutler1993,Oechslin2007,Kiuchi2010}. Since the phase of the waveform depends sensitively on the binary parameters, it is imperative that we have accurate theoretical templates in order to extract useful information from observed inspirals.

The evolution of a compact binary through radiation reaction is understood up to high order general relativistic effects in the post-Newtonian expansion, which accounts fully for the inspiral of pairs of black holes. When the binary hosts at least one neutron star, tidal interactions in principle also play a role. Tidal dissipation allows for the transfer of orbital energy into oscillations and the internal heating of the stars, which corrects the predicted rate of inspiral due to purely gravitational effects. Studies of the effect of tidal interactions and the excitation of linear perturbations have shown that tidal effects have a negligible impact on the last stages of binary inspiral~\citep{Kochanek1992,Bildsten1992,LaiRasio1994a,LaiRasio1994b,Lai1994}. In particular, they can be ignored for the purpose of gravitational wave detection and parameter extraction.

Recently, attention has been drawn to nonlinear tidal effects in close binary systems. 
\citet{Weinberg2012} investigated a variety of scenarios in which nonlinear instabilities can produce strong tidal effects and corresponding dissipation, through both the familiar parametric resonance mechanism and the less-familiar, nonlinear driving of modes due to strong mode-mode coupling. 
Even more recently, \citet{Weinberg2013} (hereafter abbreviated WAB) considered nonlinear coupling between modes in a tidally perturbed neutron star and found a potential non-resonant instability. 
The essential idea is that the tidal perturbation can set up a strong coupling between a high-order $p$-mode and a high-order $g$-mode, through a three-mode interaction term. 
Since these two daughter modes have widely spaced frequencies, with $\omega_p \gg \omega_g$, they cannot suffer from a resonant instability. But, when they have nearly identical wave numbers, $k_p \simeq k_g$, WAB found that the three-mode coupling was so strong that it destabilized the daughter modes. In this case, the tidal forces on the star rapidly drive the $g$-mode to large amplitudes, and nonlinear dissipation of these modes in turn converts the orbital energy of the binary into tidal heating of the star. Depending on the saturation amplitude, such behavior can lead to a large correction to the orbital phase of binary inspiral, at around the time that the inspiral enters into the sensitive frequency band of gravitational wave detectors. This would represent a potential difficulty for gravitational wave detection via matched filtering with a template that accumulates signal-to-noise ratio over many orbits.

In fact, the nature of the instability discussed in WAB implies that a neutron star immersed in a static tidal field is also unstable, even when the tidal perturbation is weak compared to the star's self-gravity. In this case, the star is unstable to a sort of buckling effect: the static $p$-mode would cause the star to separate radially into alternating layers of increased and decreased compression, while the static $g$-mode gives these layers an alternating horizontal shear relative to their initial positions. We may thus consider WAB's instability to be ``quasi-static'' in the sense that it exists even as the tidal forcing frequency is taken to zero (see WAB \S2 and Appendix A). However, the work of WAB focused in detail only on the three-mode coupling terms, neglecting other potentially important effects such as four-mode coupling terms.

In this paper, we present an investigation of a static, tidally perturbed star, including all of the necessary three- and four-mode terms to determine whether the star is stable to the first nonlinear corrections to perturbation theory.  In order to complete the analysis, we present a novel technique for computing the four-mode coupling terms between two daughter modes and the tidal perturbation. We find that the four-mode coupling terms cancel the three-mode coupling terms when the latter become large, protecting the star from non-resonant instabilities. We consider a non-rotating neutron star and a static tidal field in order to simplify the analysis. It is important to note that for the case of possible non-resonant instabilities, having a fixed rather than a slowly-varying tidal field (as compared to the neutron star's dynamical time scale) does not change the essential problem. This is because a quasi-static instability (such as WAB) occurs when perturbations of the deformed star can possess negative potential energy, as opposed to a parametric resonance instability where a time-varying tidal field excites oscillatory modes of positive energy (a phenomenon that is impossible if the forcing frequency vanishes). WAB also investigated parametric resonances in neutron star binaries, and found that these did not contribute significantly to the orbital evolution of the binary. Other possible effects of the time-varying tidal field are considered in Appendix~\ref{app:Rotation}, and in the discussion.

For simplicity, we only consider inviscid, normal fluid neutron stars in Newtonian gravity: this physics is sufficient to capture the instability in WAB. Including the solid neutron star crust would produce additional modes at the crust-core interface ($i$) and due to shear waves ($s$) in the crust \citep{1988ApJ...325..725M}; linear resonant excitation of the $i$-mode has been studied in the context of an energy source for gamma-ray burst precursors \citep{2012PhRvL.108a1102T}. However the quasi-static instability in WAB occurs due to mode overlap in the core, and we would not expect crustal modes to play a role. General relativistic effects are also not considered: they make modest (order $GM/R_*c^2$) perturbations to the mode frequencies, but their only qualitative effect is a small damping due to gravitational wave emission not present in the Newtonian theory \citep[see e.g.][]{1967ApJ...149..591T, 1969ApJ...158....1T, 1969ApJ...158..997T, 1983ApJ...268..837M}. We also make use of the Cowling approximation, where the background gravitational field is held fixed while the fluid elements are perturbed about their original configuration \citep[e.g.][]{Cox}, and this approximation does have a potentially important impact, since it is necessary in the technique we use to compute the four-mode coupling. In fact, the Cowling approximation is at its worst when treating the tidal deformation. However, the high-order daughter modes whose stability we are ultimately interested in should be very well described by the Cowling approximation. 

Before entering into a detailed discussion of our results, it is worthwhile to first examine a simple toy problem in order to gain an intuition regarding what order in perturbation theory we need to go to. This directly illustrates why four-mode terms are significant in the stability analysis.

\subsection{A toy model: Two dimensional oscillator}
\label{ss:toy}

Consider a two-dimensional harmonic oscillator with characteristic frequencies $\omega_1$ and $\omega_2$, such that $\omega_1 \gg \omega_2$. The potential energy of this system for a general displacement $\eta$ from equilibrium, written in coordinates with a unit-mass normalized kinetic energy term ${\mathcal T}=\frac12\dot\eta^\top\dot\eta$, is given by 
\begin{align}
\mathcal{V}(\eta) = \frac{1}{2} \eta^\top \mathcal M \eta \,, {\rm~~where~~}
\mathcal M =  \begin{pmatrix}
		\omega_1^2 & 0 \\
		0 & \omega_2^2
	\end{pmatrix}\,.
\end{align}

\begin{figure*}
\centering
\includegraphics{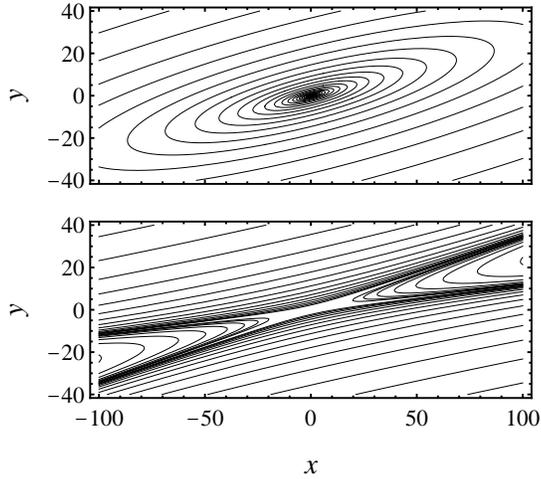}
\caption{Potential surfaces for a perturbed two-dimensional oscillator, with the perturbation taking the form of a rotation. This plot is for characteristic frequencies $\omega_1/\omega_2 = 5$ and the angle of rotation $\theta = 0.25$. The top panel shows the contours of constant potential after adding perturbations of all orders in $\theta$, while the bottom panel shows the contours after adding only terms that are first order in $\theta$. }
\label{fig:Coupledoscillator}
\end{figure*}

Consider some effect (like interaction with a third degree of freedom, for example) which rotates the coordinate basis, so that displacement vectors become $\eta^\prime = U(\theta) \eta$, where $U(\theta)$ is in the $SO(2)$ representation of the rotation. In the new basis, the potential energy is given by
\begin{align}
\mathcal{V}(\eta^\prime) =  \frac{1}{2} (\eta^\prime)^\top \mathcal M^\prime \eta^\prime \,, {\rm~~where~~}
\mathcal M^\prime =  U \mathcal M U^\top\,.
\end{align}
Putting in the form of $U(\theta)$, we see that
\begin{align}
\mathcal M^\prime (\theta) = & \begin{pmatrix}
				      \omega_1^2 \cos^2{\theta} + \omega_2^2 \sin^2{\theta} & ( \omega_1^2 - \omega_2^2 ) \cos{\theta} \sin{\theta} \\
				      ( \omega_1^2 - \omega_2^2 ) \cos{\theta} \sin{\theta} & \omega_1^2 \sin^2{\theta} + \omega_2^2 \cos^2{\theta}
				  \end{pmatrix} 
		            = \mathcal M + \delta\mathcal M, {\rm~~where}\\ 
\delta\mathcal M =&
		            \begin{pmatrix}
		            	      ( \omega_2^2 - \omega_1^2 ) \theta^2 & ( \omega_1^2 - \omega_2^2 ) \theta  \\
			               ( \omega_1^2 - \omega_2^2 ) \theta  & (\omega_1^2 - \omega_2^2) \theta^2
			          \end{pmatrix} + {\cal O}(\theta^3)\,.
\end{align}
The change in the smaller eigenvalue of $\mathcal M$ due to this change can be formally calculated using second-order perturbation theory as
\begin{align}
\omega_-^2 = & \, 
\omega_2^2 + \delta\mathcal M_{22} + \frac{\delta\mathcal M_{12}^2}{\omega_2^2 - \omega_1^2} + ...
= \omega_2^2 + \theta^2 (\omega_1^2 - \omega_2^2) + \frac{\left[( \omega_1^2 - \omega_2^2 ) \theta\right]^2}{\omega_2^2 - \omega_1^2} + \hdots = \omega_2^2\,.
\end{align}
We could have predicted this from the fact that the interactions act as a pure rotation, but this avenue of analysis highlights a fact which is useful when we do not have such global, non-perturbative information. When calculating perturbations to the eigenvalues, perturbations to the matrix elements along the diagonal ($\mathcal M_{22}$) enter into the analysis at the same order as perturbations to those off the diagonal ($\mathcal M_{12}$) squared. Hence, if the angle $\theta$ is a small angle such that $\omega_2^2/(\omega_1^2 - \omega_2^2) \simeq \omega_2^2/\omega_1^2 < \theta^2 \ll 1$, the diagonal perturbation $\delta \mathcal M_{22}$ is much smaller than the off-diagonal perturbation $\delta \mathcal M_{12}$, but the latter cannot be ignored in the calculation of $\omega_-$. Ignoring it would lead us to the conclusion that the deformed potential has an unstable direction, i.e. $ \omega_-^2 = \omega_2^2 + \theta^2 (\omega_2^2 - \omega_1^2) + \hdots < 0 $. 
Figure~\ref{fig:Coupledoscillator} illustrates this point by plotting contours of constant potential $\mathcal V$ using both the full rotation, and using only the leading order term in small angle $\theta$. In the former case, it is clear that the origin remains a stable equilibrium, but in the latter case, neglecting the higher order terms leads the origin to become a saddle point.

This example captures much of the physics describing the system we are interested in -- a tidally-deformed neutron star. In \S\ref{sec:ModeStability}, we show that in the sub-matrix of a pair of high order modes, the off-diagonal perturbation to the potential is given by the three-mode coupling between the modes and the tidal deformation, and the diagonal perturbations are given by appropriate four-mode couplings. Lessons learned from the toy model tell us that we need to evaluate both the three- and the four-mode couplings in order to determine the lowest order or $\epsilon^2$ (in dimensionless tidal strength $\epsilon$) perturbation to the eigenfrequencies. 

The example also captures one other aspect of our analysis -- what happens if the shallow direction of the potential changes with time. In the limit of $\omega_1\rightarrow\infty$, this example corresponds to that of a bead in a restoring force ($-\omega_2^2x$) sliding on a wire at position angle $\theta$. If $\theta$ varies, then the bead experiences a centrifugal or anti-restoring force ($\dot\theta^2x$). In Appendix~\ref{app:Rotation} we consider the consequences of this effect when the tidal field is varying with time due to the motion of the binary.

\subsection{Overview of the paper}

The remainder of this paper is structured as follows. We review the theory of nonlinear perturbations of a star using variational techniques in \S\ref{sec:Nonlinear}. We first discuss the Lagrangian formulation of the dynamics in \S\ref{sec:Lagrangian}, review the expansion of the problem in the basis of the linear modes of the star in \S\ref{sec:ModeExpansion}, and discuss the equilibrium tidal deformation in \S\ref{sec:StaticTide}. We arrive at the full expression of the perturbations to the mode frequencies due to three- and four-mode coupling in \S\ref{sec:ModeStability}. In \S\ref{sec:FourMode} we compute the four-mode coupling terms using a novel technique. We describe this technique in \S\S\ref{sec:VTransform}--\ref{sec:Understanding}, apply it to recast the problem of perturbations of the mode frequencies in \S\S\ref{sec:PotentialB}--\ref{sec:Matching}, and show that largest potentially unstable terms cancel in \S\ref{sec:ModeStabilityRevisited}. In \S\ref{sec:Compute} we estimate the size of the remaining perturbations, and show that they are small in the case of high-order daughter modes in the presence of the tidal deformation. Finally, we conclude with an overview and discussion of our results in \S\ref{sec:Discussion}. Some technical details that arise along the way are collected into the Appendices.

\section{Perturbations in tidally deformed stars}
\label{sec:Nonlinear}

We now discuss the equations which govern stellar perturbations, focusing on three- and four-mode interactions between tidal deformations and additional perturbations of the star. First we review the Lagrangian formulation of general perturbations of a background star, and then we expand these perturbations in the basis of the linear modes of the star. This allows us to examine how nonlinear interactions perturb the eigenfrequencies of the linear perturbations. We show how coupling between modes can generate an instability, and we also show that in order to determine if an instability exists, we must account for both three- and four-mode interactions.

\subsection{Lagrangian formulation of perturbations}
\label{sec:Lagrangian}

Consider a non-rotating, inviscid, fluid neutron star with total mass $M$, radius $R_*$, and a characteristic dynamical frequency $\omega_0^2 = G M/R_*^3$, which is perturbed by a weak tidal field. The tidal potential has the form $\epsilon U({\bm x})$, where $\epsilon$ is the dimensionless tidal strength. Our particular expression for $U$ is the leading-order tidal potential due to a distant companion star of mass $m$, held at a fixed separation $a$, so that $\epsilon =  G m/(\omega_0^2 a^3)$. We consider the leading multipole of the corresponding tidal potential, which is
\begin{align}
\label{eq:Tide}
U = - \omega_0^2 r^2 P_2 (\cos \theta) \,,
\end{align}
where $P_l$ is a Legendre polynomial with unit normalization, and $\theta$ measures the angle from the line joining the star to its companion. The perturbing potential is axisymmetric in this coordinate system. This work can be generalized to include the higher, weaker multipoles of the potential in a straightforward manner.

Let $\bm{\xi}$ denote a general displacement field on the star. Each fluid element of the star is labeled by its initial coordinates, ${\bm x}$ (the Lagrangian coordinates of the element), so that elements initially at ${\bm x}$ are displaced to their actual (Eulerian) positions ${\bm x}'$ as ${\bm x} \to {\bm x}' = {\bm x} + {\bm \xi}$. Whatever coordinate system we choose to use, whether it be ${\bm x}$, ${\bm x}'$, or some other coordinate system ${\bm X}$, note that the masses of the elements are invariant, so that $d^3 x \rho({\bm x}) = d^3 x' \rho'({\bm x}') = d^3 X \rho_{\bm X} ( {\bm X})$. From now on, quantities such as $\rho$ are taken to be defined in the coordinate system being used at that stage.

The displacement field has a Lagrangian $\mathcal L$ given by
\begin{equation}
\label{eq:GeneralL}
\mathcal{L}({\bm \xi},\dot{\bm \xi})=\int d^3 x \,  \rho({\bm x}) \, \frac{1}{2} \dot{\bm \xi}({\bm x})\cdot \dot{\bm \xi}({\bm x})  - \mathcal V ({\bm \xi}) \,,
\end{equation}
where the potential $\mathcal V(\xi)$ incorporates both the internal energy of the perturbed star and the gravitational energy. For a star with an external perturbing field, the potential energy with an arbitrary displacement field takes the form
\begin{align}
\label{eq:GeneralV}
\mathcal V({\bm \xi}) = \int d^3 x' \, \rho({\bm x}') & \left[ \mathcal E_{\rm int} ( {\bm x}') + \Phi_0 ( {\bm x}') + \epsilon U ( {\bm x}') \right] + \mathcal C \notag \\
= \int d^3 x \, \rho({\bm x}) & \left[ \mathcal E_{\rm int} ( {\bm x} + {\bm \xi}) + \Phi_0 ( {\bm x} + {\bm \xi}) + \epsilon U ( {\bm x} + {\bm \xi}) \right] + \mathcal C \notag \\
= \int d^3 x \, \rho({\bm x}) &\left[ \mathcal E_{\rm int} ( {\bm x}) + \Phi_0 ({\bm x})  + \frac 12 {\bm \xi} \cdot {\bm C} \cdot {\bm \xi} + \frac{1}{3!}f_3({\bm \xi},{\bm \xi},{\bm \xi})+\frac{1}{4!}f_4({\bm \xi},{\bm \xi},{\bm \xi},{\bm \xi})+\hdots \right. \notag \\
&\left. +\epsilon \, {\bm \xi} \cdot {\bm \nabla} U +\frac{1}{2} \epsilon\,  {\bm \xi} \cdot({\bm \xi} \cdot {\bm \nabla}) {\bm \nabla} U + \hdots \right] + \mathcal{C} \,.
\end{align}

In the case of linear perturbations, only the symmetric bilinear ${\bm C}$ and the gradient of $U$ contribute to the dynamics of the displacement field. Physically, ${\bm C}\cdot{\bm \xi}$ is the linear restoring force opposing an infinitesimal displacement ${\bm \xi}$. \citet{LyndenBell1967a} \citep[see also, e.g.][]{Schenk2002} derive a functional form for ${\bm C}$. The functionals $f_n$, meanwhile, encode the nonlinear corrections to the restoring forces due to the internal and self-gravitational energy of the star. These functionals are symmetric and linear under addition and scalar multiplication of displacements, but not under multiplication of the displacements by scalar functions. Later, when we expand the displacement in terms of the linear modes, they give us the $n$-mode couplings.

The term $\mathcal C$ in this case contains the contributions to the energy from the gravitational field itself, which are fixed by the Cowling approximation and do not contribute to the dynamics of ${\bm \xi}$. The issue of the appropriate division of gravitational potential energy between the field degrees of freedom and the interaction between the fluid elements and those fields is discussed in Appendix~\ref{sec:FieldDOF}. We also remark that for the tidal field given in Eq.~\eqref{eq:Tide}, $\int d^3 x \, \rho \, U = 0$  due to the fact that the background $\rho$ is spherically symmetric and $U$ is dipolar, and so has not been written in Eq.~\eqref{eq:GeneralV}. In addition, since $U$ is a quadratic function of ${\bm x}$, all third and higher derivatives of $U$ vanish, and as such have not been written.

\subsection{Mode expansion of the Lagrangian}
\label{sec:ModeExpansion}

The most convenient language in which to discuss nonlinear perturbations is in terms of an expansion in the orthonormal basis of linear modes of the star. We write the mode functions themselves are as ${\bm \xi}_a$, and a general displacement can be expanded as
\begin{align}
{\bm \xi} ({\bm x}) = \sum_a c_a {\bm \xi}_a ({\bm x}) 
\end{align}
using this basis. In spherical coordinates, the mode functions have the form\footnote{We do not consider toroidal displacements, ${\bm \xi} \propto {\bm r} \times {\bm \nabla} Y_{a}$, since they do not couple to the tidal field~\citep[e.g.][]{Cox}. }
\begin{align}
\label{eq:chiexpansion}
{\bm \xi}_a = \xi_r Y_a \hat{\bm r} + \xi_h \left ( \partial_\theta Y_a \hat{\bm \theta} + \frac{1}{\sin{\theta}} \partial_\phi Y_a \hat{\bm \phi} \right) \,,
\end{align}
where $\xi_r$ and $\xi_h$ are functions of $r$ and $Y_a = Y_{l_a,m_a} (\theta,\phi)$. The basis obeys the orthonormality condition
\begin{align}
\int d^3 x \, \rho\, {\bm \xi}_a^* \cdot  {\bm \xi}_b = \frac{E_0}{\omega_a^2} \delta_{ab} \,,
\end{align}
where $E_0 = G M^2/R_*$. This normalization, which is the same as the one used by WAB but differs from that of \citet{Schenk2002}, means that when a basis mode is excited to unit amplitude, it has energy $E_0$. By expanding in these basis functions, we can re-express our Lagrangian and potential as sums over mode amplitudes.

The square of the mode frequencies $\omega_a^2$ are the eigenvalues of the bilinear ${\bm C}$ in the mode basis,
\begin{align}
{\bm C} \cdot {\bm \xi}_a = \omega_a^2  {\bm \xi}_a\,.
\end{align}
By defining the mode expansions of the remaining terms in the Lagrangian, we can write it entirely in terms of the mode amplitudes. The definitions we need are
\begin{align}
\label{eq:UaMode}
U_a =& - \frac{1}{E_0} \int d^3 x \, \rho \, {\bm \xi}_a^* \cdot {\bm \nabla } U \,, \\
\label{eq:UabMode}
U_{ab} =& - \frac{1}{E_0} \int d^3 x \, \rho \, {\bm \xi}_a \cdot ({\bm \xi}_b \cdot{\bm \nabla }){\bm \nabla } U \,, \\
\label{eq:kappa3}
\kappa_{abc} =& - \frac{1}{2E_0}  \int d^3 x \, \rho \, f_3({\bm \xi}_a,{\bm \xi}_b,{\bm \xi}_c) \,, {\rm~~and}\\
\label{eq:kappa4}
\kappa_{abcd} =& - \frac{1}{3! E_0}  \int d^3 x \, \rho \, f_4({\bm \xi}_a,{\bm \xi}_b,{\bm \xi}_c,{\bm \xi}_d) \,.
\end{align}
Due to the symmetry of the functionals $f_n$, the coupling terms $\kappa_{abc}$ and $\kappa_{abcd}$ are symmetric in all their indices. 
When we expand the Lagrangian in this mode basis, we use the fact that the displacement ${\bm \xi}$ is real, and can be replaced by ${\bm \xi}^{*}$ as and when needed. Using these definitions and the potential~\eqref{eq:GeneralV}, we rewrite the Lagrangian~\eqref{eq:GeneralL} in the form
\begin{align}
\label{eq:ModeL}
\mathcal L = & E_0 \sum   \left[ \frac {\dot{c}_a \dot{c}_a^{*}}{2 \omega_a^2} - \frac 12 c_a c_a^{*} + \frac13 \kappa_{\bar{a} \bar{b} \bar{c}} c_a^{*} c_b^{*} c_c^{*} + \frac 14 \kappa_{\bar{a} \bar{b} \bar{c} \bar{d}} c_a^{*} c_b^{*} c_c^{*} c_d^{*} + \epsilon U_a c_a^{*} + \frac12 \epsilon U_{\bar{a} \bar{b}} c_a^{*} c_b^{*} + \hdots \right] - \mathcal V(0) \,.
\end{align}
Here, our convention is that the sum runs over all the repeated indices in each term (including the index $a$ in the first two terms). 
Bars indicate the use of complex conjugate wave-functions in the respective terms. Note that in the first and second terms, the amplitude of a mode with a given $m$ comes up twice -- in the terms corresponding to $m$ and in those corresponding to $-m$, since the reality of ${\bm \xi}$ implies that they are related by
$c_{a_{-m}}=(-1)^m c_{a_m}^\ast$. 
We recall that the potential term $\mathcal V$ with the displacement set to zero has the form
\begin{align}
\mathcal V(0) = \int d^3 x \, \rho   &\left[ \mathcal E_{\rm int} ( {\bm x}) + \Phi_0 ({\bm x}) \right] + \mathcal C \,.
\end{align}
We use the rule that sums run over repeated indices throughout this work, except where noted. Though we do not do so here, the equations of motion, as presented in e.g. \citet{Schenk2002}, can be derived by varying this Lagrangian with respect to the amplitude $c_a^*$.

With the formalism of the mode expansion established, we now investigate the perturbations excited by a static tidal field.

\subsection{The static response to the tide}
\label{sec:StaticTide}

The star responds to the external tidal field $\epsilon U$ by deforming to a new equilibrium configuration. We denote this equilibrium tidal deformation ${\bm \chi}$ and set ${\bm \xi} $ equal to ${\bm \chi}$ in the preceding equations. The tidal displacement ${\bm \chi}$ is such the internal restoring forces balance the external perturbing force,
\begin{align}
\left. \frac{\delta \mathcal V}{\delta {\bm \xi}}\right|_{{\bm \xi} \to {\bm \chi}} = 0 \,.
\end{align}
Physically, ${\bm \chi}$ takes the spherically symmetric star to a static configuration where elements along contours of constant gravitational potential $\Phi_0 + \epsilon U$ have the same density and pressure. The displacement ${\bm \chi}$ has a part linear in the tidal strength $\epsilon$, which is the so-called linear tide. There are also higher terms, which arise from the nonlinear restoring forces and the nonlinear tide. In order to consistently account for the changes in the eigenfrequencies $\omega_a$ of additional perturbations due to coupling with the tide, we need to keep terms up to $O(\epsilon^2)$. As such, we write the expansion of ${\bm \chi}$ in the mode basis as
\begin{align}
\label{eq:tidaldisp}
{\bm \chi} =\sum \chi_a {\bm \xi}_a = \sum (\epsilon \chi^{(1)}_a + \epsilon^2 \chi^{(2)}_a) {\bm \xi}_a \,.
\end{align}

Varying just the potential terms of the Lagrangian~\eqref{eq:ModeL} with respect to a given mode amplitude $c_a^*$, and then substituting the amplitudes $\chi_a$ into the result gives an equation for $\chi_a$,
\begin{align}
  \chi_a- \epsilon U_a - \sum\left[\kappa_{\bar{a} \bar{b} \bar{c}} \chi_b^{*}\chi_c^{*} + \kappa_{\bar{a} \bar{b} \bar{c} \bar{d}} \chi_b^{*}\chi_c^{*}\chi_d^{*} + \epsilon U_{\bar{a} \bar{b}} \chi_b^{*} +\hdots \right]&=0 \,.
\end{align}
Solving order by order for $\chi_a$, we find that
\begin{align}
\label{eq:chi1}
\chi^{(1)}_a&=U_a {\rm~~and}\\
\label{eq:chi2}
\chi^{(2)}_a&=\sum \left[ \kappa_{\bar{a} \bar{b} \bar{c}} U_b^{*} U_c^{*} + U_{\bar{a} \bar{b}} U_b^{*} \right] \,.
\end{align}
Given the tidal field $\epsilon U$, the fluid elements respond by the nonlinear displacement ${\bm \chi}$. Note that since $U$ is axisymmetric, ${\bm \chi}$ is also axisymmetric and $\chi_a$ contains only terms where $m=0$. This means that all of the terms in Eqs~\eqref{eq:chi1} and~\eqref{eq:chi2} are actually real. The question we must answer is whether or not this deformed star is stable. 

\subsection{Stability of the deformed star}
\label{sec:ModeStability}

Now we allow the star to undergo further perturbations, so that 
\begin{align}
\label{eq:FullDisplace}
{\bm \xi} = {\bm \chi} + {\bm \eta}\,,
\end{align}
where ${\bm \eta}$ is an additional displacement field on the star away from the equilibrium configuration. We are interested in the interaction of pairs of daughter modes with the tidal perturbation, and so we expand our Lagrangian only to second order in the additional small perturbations, $O(\eta^2)$. We do not deal with the coupling of three or more non-tidal modes either to each other or to the tide, and so $O(\eta^3)$ and higher terms are dropped in what follows.

By expanding ${\bm \eta}$ as $\sum \eta_a {\bm \xi}_a$, and substituting the corresponding expansion of the displacement $c_a = \chi_a + \eta_a$ into the Lagrangian~\eqref{eq:ModeL}, we arrive at the following form of the Lagrangian, up to the order we are interested in,
\begin{align}
  \mathcal L =&  E_0 \sum \left[ \frac{\dot{\eta}_a \dot{\eta}_a^{*}}{2 \omega_a^2} - \frac12 (\chi_a \chi_a^{*} + \eta_a \eta_a^{*} ) + \kappa_{\bar{a} \bar{b} \bar{c}} \eta_a^{*} \eta_b^{*} \chi_c^{*} + \frac 32 \kappa_{\bar{a} \bar{b} \bar{c} \bar{d}} \eta_a^{*} \eta_b^{*} \chi_c^{*} \chi_d^{*} + \epsilon U_a \chi_a^{*} + \frac 12 \epsilon U_{\bar{a} \bar{b}} \eta_a^{*} \eta_b^{*} \right] - \mathcal V(0) \nonumber\\
\label{eq:etaL}
=& E_0 \sum \left[ \frac{\dot{\eta}_a \dot{\eta}_a^{*}}{2 \omega_a^2} - \frac12 \eta_a \eta_a^{*} + \left(  \epsilon [\kappa_{\bar{a} \bar{b} \bar{c}} \chi_c^{(1)*} + \frac12 U_{\bar{a} \bar{b}}] + \epsilon^2 [\kappa_{\bar{a} \bar{b} \bar{c}} \chi_c^{(2)*} + \frac{3}{2} \kappa_{\bar{a} \bar{b} \bar{c} \bar{d}} \chi_c^{(1)*} \chi_d^{(1)*} ] \right) \eta_a^{*} \eta_b^{*} \right] + \frac {E_0}{2 }\sum \epsilon^2 |U_a|^2 - \mathcal V(0) \,.
\end{align}
In the first line we have written only those terms which are of the correct order, and we have also neglected terms which have only a single factor of $\eta_a$, since these vanish upon substitution of our solution for $\chi_a$. The terms linear in ${\bm \eta}$ vanish because the displacement ${\bm \chi}$ takes the star to an equilibrium state. In the next line, we have substituted the decomposition~\eqref{eq:tidaldisp}, collected terms in orders of $\epsilon$, and used Eq.~\eqref{eq:chi1} to isolate the $|U_a|^2$ term. For now, we ignore the overall constant terms $\mathcal V(0)$ and $E_0 \sum \epsilon^2 |U_a|^2/2$, which follow along for the ride but do not affect the dynamics.

The $O(\epsilon)$ term in Eq.~\eqref{eq:etaL} is the three-mode interaction $\kappa_{\bar a \bar b \bar c} \eta_a^* \eta_b^* \chi_c^{(1)*}$. If the coupling $\kappa_{\bar p \bar g \bar c} \chi_c^{(1)*}$ happens to be large and positive for a particular pair of modes $(p,g)$ then it can overcome the smallness of the tidal coupling strength $\epsilon$. As first noted by WAB, $\kappa_{\bar p \bar g \bar c} \chi_c^{(1)*}$ is in fact large for certain high order $p$-mode and $g$-mode pairs, due to a spatial resonance of the mode functions ${\bm \xi}_p$ and ${\bm \xi}_g$. However, as we show below, this three-mode term perturbs the characteristic frequencies at $O(\epsilon^2)$, as do all of the terms in Eq.~\eqref{eq:etaL} multiplied by $\epsilon^2$. It is not immediately clear what role these terms play.

By varying the Lagrangian~\eqref{eq:etaL} with respect to the amplitudes $\eta_a$ we can derive the equations of motion, and from there the characteristic frequencies. We instead adopt an equivalent, but hopefully more transparent strategy to obtain the characteristic frequencies. Defining rescaled amplitudes $\eta_a ' = \eta_a/\omega_a$ leads us to the analogy to a system of coupled oscillators, for which we can rewrite the Lagrangian in matrix notation as
\begin{align}
\mathcal L = \frac {E_0}{2} (\dot{\eta}' )^\dagger \dot{\eta}' - \frac {E_0}{2} (\eta')^\dagger \mathcal M \eta' \,,
\label{eq:MatrixL}
\end{align}
where $\eta'$ is a vector of mode amplitudes and the matrix $\mathcal M$ contains the leading restoring terms, the effect of the tidal potential, and the mode-mode interactions. The rescaling introduces factors of $\omega_a$ at each instance of an $\eta_a$, and it is equivalent to normalizing the mode functions to all have the same moment of inertia $M R_*^2$ at unit amplitude rather than the same energy $E_0$, which is the choice of normalization used in \cite{Schenk2002}. 

The eigenvalues of $\mathcal M $ are the characteristic frequencies of this set of oscillators, and consideration of $\mathcal M$ is equivalent to writing out the equations of motion and solving the corresponding characteristic equation. Because the tidal potential is axisymmetric, there is an ordering of the basis modes where the matrix $\mathcal M$ is made up of $2\times2$ and $1\times1$ blocks along the diagonal, where each block consists of those modes which are allowed to interact given conservation of angular momentum. The blocks are all independent of each other, and analysis of one essentially applies to the rest. 

We now focus on a particular pair of modes with $m=0$, which we give indices $p$ and $g$, and write the sub-block of $\mathcal M$ for this pair. We discuss the structure of $\mathcal M$ and the case of modes where $m\neq 0$ further in Appendix~\ref{sec:Nonaxisymm}. The modes have unperturbed frequencies $\omega_p$ and $\omega_g$, and we set $\omega_p > \omega_g$. The sub-block of $\mathcal M$ is composed of
\begin{align}
\label{eq:masspp}
\mathcal M_{pp} = & \omega_p^2 - \epsilon \omega_p^2 \left( U_{pp} + \sum 2 \kappa_{app} \chi_a^{(1)} \right) - \epsilon^2 \omega_p^2 \sum\left[ 2 \kappa_{app} \chi_a^{(2)} + 3 \kappa_{abpp} \chi_a^{(1)} \chi_b^{(1)} \right] \,, \\
\label{eq:masspg}
\mathcal M_{pg} = & \mathcal M_{gp} = -\epsilon \,  \omega_p \omega_g \left(U_{pg} + \sum 2 \kappa_{apg} \chi_a^{(1)} \right) \,, {\rm~~and}\\
\label{eq:massgg}
\mathcal M_{gg} = & \omega_g^2 - \epsilon \omega_g^2 \left( U_{gg} + \sum 2 \kappa_{agg} \chi_a^{(1)} \right) - \epsilon^2 \omega_g^2 \sum\left[ 2 \kappa_{agg} \chi_a^{(2)} + 3 \kappa_{abgg} \chi_a^{(1)} \chi_b^{(1)} \right] \,,
\end{align} 
and note that since $p$ and $g$ denote particular modes, they are not summed over in the expressions for $\mathcal M$. Formally expanding in powers of $\epsilon$, we can compute the perturbed eigenvalues of this matrix as
\begin{align}
\label{eq:omegaplus}
  \omega_+^2 = &\, \omega_p^2 - \epsilon \omega_p^2 \left( U_{pp} + \sum 2 \kappa_{app} \chi_a^{(1)} \right) 
  - \epsilon^2 \omega_p^2 \sum\left[2 \kappa_{app} \chi_a^{(2)} + 3 \kappa_{abpp} \chi_a^{(1)} \chi_b^{(1)} \right]
  + \epsilon^2 \frac{ \omega_p^2 \omega_g^2 }{\omega_p^2 - \omega_g^2}\left( U_{pg} + \sum 2 \kappa_{apg} \chi_a^{(1)} \right)^2 
\end{align}
and
\begin{align}
\label{eq:omegaminus}
  \omega_-^2 = &\, \omega_g^2 - \epsilon \omega_g^2 \left( U_{gg} + \sum 2 \kappa_{agg} \chi_a^{(1)} \right) 
  - \epsilon^2 \omega_g^2 \sum\left[2 \kappa_{agg} \chi_a^{(2)} + 3 \kappa_{abgg} \chi_a^{(1)} \chi_b^{(1)} \right]
  - \epsilon^2 \frac{ \omega_p^2 \omega_g^2 }{\omega_p^2 - \omega_g^2}\left( U_{pg} + \sum 2 \kappa_{apg} \chi_a^{(1)} \right)^2 .
\end{align}
Since $\omega_g$ is the smaller frequency to begin with, the negative-definite term involving $\kappa_{apg}$ in Eq.~\eqref{eq:omegaminus} is in danger of pushing $\omega_g$ to a negative value, especially if the mode is a high order $g$-mode with $\omega_g^2 \ll \omega_0^2 = G M/R_*^3 \ll \omega_p^2$. This is the potential nonlinear instability described by WAB.

In order to make definite statements, we need to know how the other terms behave, in particular the four-mode interaction term $\kappa_{abgg} \chi_a^{(1)} \chi_b^{(1)}$. If they are large when the three-mode coupling is large, they can in principle prevent the instability. We now discuss a method for computing the interaction between two daughter modes and the tidal deformation at the level of the four-mode coupling. Along the way we also find a way to simply derive the three-mode terms. We find that the four-mode term serves to precisely cancel the $\kappa_{apg}$ term.

\section{Computing the four-mode coupling terms}
\label{sec:FourMode}

\begin{figure*}[t]
\centering
\includegraphics[width=3.2in]{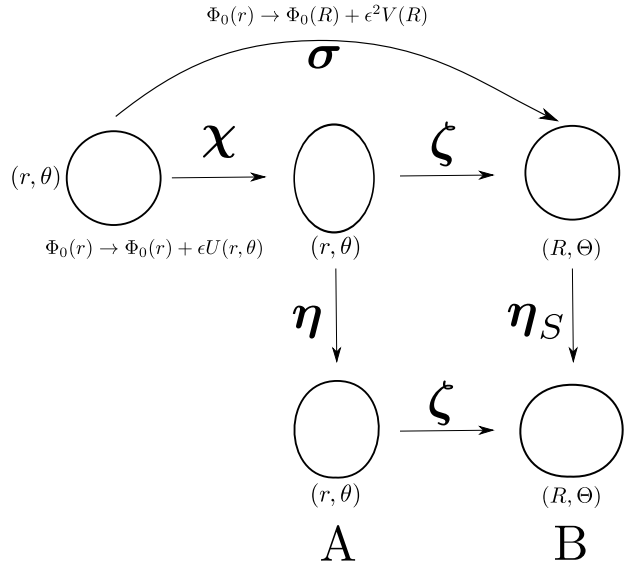}
\caption{Schematic of the two stellar models. The external tidal field produces a deformation ${\bm \chi}$. In star A, we consider further perturbations ${\bm \eta}$. Star B is obtained by mapping the tidally deformed star back to a spherically symmetric configuration by a coordinate transform. Further perturbations are now expressed in this new coordinate system; the field ${\bm \eta}_S$ is that perturbation in model B which corresponds to the perturbation ${\bm \eta}$ in model A. }
\label{fig:Models}
\end{figure*}

\begin{table}
\centering
\caption{Key for the vector quantities involved in the two stellar models
\label{tab:Definitions}}
\begin{tabular}{c c}
Vector Quantity & Definition \\
\tableline
${\bm \chi}$ & Equilibrium tidal deformation \\
${\bm \zeta} $& Infinitesimal generator of the volume-preserving coordinate transform $\psi$ \\
${\bm \sigma}$ & Combined application of the equilibrium tide and the coordinate transform $\psi$ \\
${\bm \eta}$ & Further, general perturbation of Star A \\
${\bm \eta}_S$ & Further, general perturbation of Star B \\
${\bm 1},{\bm 2}, {\bm 3}$ & Virtual displacement vectors that sum to $\bm \chi$ \\
${\bm J}_\psi$ & The Jacobian transformation matrix for the coordinate transform $\psi$\\
\tableline\\ \\
\end{tabular}
\end{table}

In this section we derive the four-mode terms discussed in \S\ref{sec:ModeStability}. The method we use is straightforward in practice, but it is not obvious from the outset that such a method will be useful. Briefly, we utilize a coordinate transform that removes the lowest order perturbation, thereby restructuring the orders of $\epsilon$ in the potential of the star. By matching the mode expansions of the potential of the transformed star with those derived in the original coordinates, we can write higher-order mode coupling terms as functions of the lower terms and the coordinate transform itself. 

Figure~\ref{fig:Models} shows the two stars -- the star in the original coordinates and in the transformed coordinates -- and the various perturbations that are applied to the stars. The first, which we call star A, has already been given a full treatment in \S\ref{sec:Nonlinear}. There we considered the interactions between the tidal displacement ${\bm \chi}$ and further perturbations ${\bm \eta}$. In star B, we use a volume-preserving coordinate transform to map the tidally deformed star back into a spherically symmetric configuration. This transform is generated by an infinitesimal, volume-preserving displacement field ${\bm \zeta}$. We then consider further perturbations on star B. It is important to note that A and B are the same stars, just in different coordinate systems. Table~\ref{tab:Definitions} provides a key for the various vector quantities which are used in the discussion of the two stars.

In order to gain any new insights from star B, we expand the potentials of both stars in the same mode basis, which is the basis of linearized modes of the original, unperturbed star. The mode functions in star B are linear combinations of the original mode functions, and so are related by the Jacobian matrix of the coordinate transform, expanded in the mode basis. We find that we can write the three- and four-mode couplings in star A as functions of this Jacobian and a modified tidal perturbation field.

\subsection{The volume-preserving transform}
\label{sec:VTransform}

To begin with, we apply the tidal perturbation $\epsilon U$ to a spherical star, so that the total gravitational potential becomes $\Phi_0 + \epsilon U$. In response the fluid elements of the star are displaced by the field ${\bm \chi}$ as discussed in \S\ref{sec:StaticTide}. Next, we consider a coordinate transform $\psi: {\bm x} = (r, \theta, \phi) \to {\bm X} = (R, \Theta, \phi)$. We require this transform to have the special properties that it is volume-preserving, and that it returns the star to spherical symmetry. These properties completely determine our mapping in terms of a power series expansion in $\epsilon$. The resulting spherically symmetric star is not the original star, because ${\bm \chi}$ itself is not volume-preserving; but as we will see the coordinate transform reverses ${\bm \chi}$ at leading order in $\epsilon$. Since our problem is axisymmetric, the $\phi$ coordinate is unchanged, and from here on we can safely ignore the coordinate $\phi$.

We require an expansion of the coordinate transform in orders in $\epsilon$, and so we represent $\psi$ by a coordinate flow under an infinitesimal generator ${\bm \zeta}({\bm x})$. Explicitly, the finite transformation $\psi$ generated by the infinitesimal transformation ${\bm\zeta}$ is defined via a parameterized transform $\varphi(s,{\bm x})$ by the ordinary differential equation $d\varphi(s,{\bm x})/ds = {\bm\zeta}(\varphi(s,{\bm x}))$, initial condition $\varphi(0,{\bm x})={\bm x}$, and assignment ${\bm X} = \psi({\bm x}) \equiv \varphi(1,{\bm x})$. The flow forward, $\psi$, gives us our new coordinates in the form of a Taylor expansion
\begin{align}
\label{eq:ForwardT}
{\bm X} = {\bm x}+ {\bm \zeta}({\bm x})+ \left. \frac 12 ({\bm \zeta} \cdot {\bm \nabla}) {\bm \zeta} \right|_{\bm x} \,,
\end{align}
up to order ${\bm\zeta}^2$. The volume-preserving requirement is equivalent to requiring the generator to be divergenceless, $\nabla\cdot{\bm \zeta}=0$.\footnote{The Jacobian $J^i{_j}(s,{\bm x}) = \partial \varphi^i(s,{\bm x})/\partial x^j$ satisfies the differential equation or chain rule $\partial J^i{_j}/\partial s = \partial^2 \varphi^i(s,{\bm x})/\partial x^j\partial s = \partial\zeta^i(\varphi(s,{\bm x}))/\partial x^j = \zeta^i{_{,k}}(\varphi(t,{\bm x}))\,J^k{_j}$, from which we see that its determinant $|{\bf J}|$ satisfies $\partial\ln|{\bf J}|/\partial s = \zeta^i{_{,i}}(\varphi(s,{\bm x}))$ -- thus a divergenceless ${\bm\zeta}$ leads to a volume-preserving transformation.} It is actually more convenient to first get an explicit representation of $\psi^{-1}$: this is actually the transformation generated by the negative of the generator. (This can be seen from the integral curve definition of a generator if we reverse the sign of $d/ds$; for an explicit proof see Appendix~\ref{sec:Reverse}.) Thus the mapping from ${\bm X}$ to ${\bm x}$ via $-{\bm\zeta}$ is
\begin{align}
\label{eq:InverseT}
{\bm x} = {\bm X} + (-{\bm \zeta})({\bm X}) +\left. \frac 12 [(-{\bm \zeta}) \cdot {\bm \nabla}](- {\bm \zeta}) \right|_{\bm X}\,.
\end{align}
Because of this, once we find an expression for the generator of the inverse flow $-{\bm \zeta}$ in $(R,\Theta)$ coordinates, we have the desired generator ${\bm \zeta}({\bm x})$.\footnote{A similar construction is used in the Lie transformation theory of small but finite canonical transformations of Hamiltonian systems \citep[e.g.][]{1969CeMec...1...12D}. There a transformation is generated by the infinitesimal flow in phase space ${\bm\zeta}({\bm x},{\bm p})$; the requirement for the transformation to be canonical corresponds to the requirement that ${\bm\zeta}$ be derivable from a Hamiltonian. The rule that a transformation can be inverted by reversing the sign of the generator is the same.}

The vector $-{\bm \zeta}({\bm X})$ has an expansion in powers of $\epsilon$, which is
\begin{align}
{\bm \zeta}   = \epsilon {\bm \zeta}^{(1)} + \epsilon^2 {\bm \zeta}^{(2)} + O (\epsilon^3) \,.
\end{align}
 From Eq.~\eqref{eq:InverseT}, the magnitude $r$ of the coordinate vector ${\bm x}$ is
\begin{align}
\label{eq:rexpand}
r = R- \epsilon  \hat{\bm R} \cdot {\bm \zeta}^{(1)}  +\epsilon^2 \left[ \frac{   {\bm \zeta}^{(1)} \cdot  {\bm \zeta}^{(1)}}{2 R} - \frac{ (\hat{\bm R} \cdot  {\bm \zeta}^{(1)})^2}{2R} + \frac{\hat{\bm R} \cdot ({\bm \zeta}^{(1)} \cdot {\bm \nabla})  {\bm \zeta}^{(1)}}{2} - \hat{\bm R} \cdot  {\bm \zeta}^{(2)}\right] \,,
\end{align}
which can be further simplified using the vector identity $ ({\bm A} \cdot {\bm \nabla}) {\bm A} = {\bm \nabla} A^2 / 2- {\bm A} \times ({\bm \nabla} \times {\bm A})$, to yield
\begin{align}
r = R - \epsilon \zeta^{(1)}_R  +\epsilon^2 \left[ \frac{ \zeta^{(1)}_R \partial_R   \zeta^{(1)}_R}{2} +  \frac{ \zeta^{(1)}_\Theta \partial_\Theta  \zeta^{(1)}_R}{2R} -  \zeta^{(2)}_R \right] \,,
\end{align}
with the definitions ${\bm \zeta}^{(i)} \cdot \hat{\bm R} =  \zeta^{(i)}_R  $ and ${\bm \zeta}^{(i)} \cdot \hat{\bm \Theta} =  \zeta^{(i)}_\Theta$. Also useful is the expansion
\begin{align}
P_2 ( \cos \theta) =  (1 - \epsilon {\bm \zeta} \cdot {\bm \nabla}) P_2 (\cos \Theta) =  P_2 (\cos \Theta) - \epsilon \frac{\zeta^{(1)}_\Theta \partial_\Theta P_2(\cos \Theta)}{R}
\end{align}
to the order needed. Now we are in place to solve for ${\bm \zeta}$, and at the same time to determine the form of the total gravitational potential $\Phi(R)$ out to order $\epsilon^2$. The tidally perturbed potential is
\begin{align}
\Phi_0(r) - \epsilon \omega_0^2 r^2 P_2 (\cos \theta) =& \Phi_0 (R) - \epsilon \left[\zeta^{(1)}_R  \mathfrak g + \omega_0^2 R^2 P_2(\cos \Theta)  \right] 
+ \epsilon^2 \left[ \left(  \frac{\zeta^{(1)}_R \partial_R   \zeta^{(1)}_R}{2} +  \frac{ \zeta^{(1)}_\Theta \partial_\Theta  \zeta^{(1)}_R}{2R} -  \zeta^{(2)}_R \right)  \mathfrak g  \right. \notag \\
& \left. + \frac 12 (\zeta^{(1)}_R)^2 \frac{d \mathfrak g}{dr} +  2 \omega_0^2 R  \zeta^{(1)}_R P_2(\cos \Theta) + \omega_0^2 R \zeta^{(1)}_\Theta \partial_\Theta P_2(\cos \Theta) \right] \,.
\end{align}
For convenience here and later, we have denoted the local gravitational acceleration as $\mathfrak g = d \Phi_0/dR$.

By insisting that all $\Theta$ dependence in $\Phi$ vanishes, we find at first order
\begin{align}
\label{eq:zeta1}
\zeta^{(1)}_R = - \omega_0^2 \frac{R^2 }{\mathfrak g}P_2 (\cos \Theta )\,,
\end{align}
and at second order
\begin{align}
\label{eq:zeta2}
\zeta^{(2)}_R & =  \frac{\zeta^{(1)}_R \partial_R   \zeta^{(1)}_R}{2} +  \frac{ \zeta^{(1)}_\Theta \partial_\Theta  \zeta^{(1)}_R}{2R} + \frac { (\zeta^{(1)}_R)^2 }{2R} \frac{d \ln \mathfrak g}{d \ln R}  + \frac{\omega_0^2}{\mathfrak g}\left[ 2 R \zeta^{(1)}_R P_2(\cos \Theta) +  R\zeta^{(1)}_\Theta \partial_\Theta P_2(\cos \Theta)\right] \,.
\end{align}
In order to write this last equation entirely in terms of the gravitational potentials, we need an expression for $\zeta_\Theta^{(1)}$. For this, we use the requirement that $\psi$ be volume-preserving. The volume-preserving infinitesimal displacements are spanned by the vectors
\begin{align}
{\bm f}^1_{lm} &= l(l+1) \frac{u_{lm}}{R} \hat{\bm R} + \left( \frac {u_{lm}}{R} + \partial_R u_{lm} \right) R {\bm \nabla} Y_{lm} \,{\rm~~and} \\
{\bm f}^2_{lm} & = w_{lm} {\bm R} \times {\bm \nabla} Y_{lm} \,.
\end{align}
This can be seen by considering the usual decomposition of vectors into vector spherical harmonics, and finding those combinations which satisfy ${\bm \nabla} \cdot {\bm f}_{lm} = 0$. These naturally decompose into the spheroidal displacements ${\bm f}_{lm}^1$ and the toroidal displacements ${\bm f}_{lm}^2$~\citep[see also][Ch. 12]{BlandfordThorne}. We are interested in axially symmetric perturbations, so we only need to use the basis vectors ${\bm f}^1_{lm}$. We further absorb the normalization of the spherical harmonics into the definition of the $u_{lm}$ and write (with $m=0$ implicit)
\begin{align}
\label{eq:zetadef}
{\bm \zeta} &= \sum_l l (l+1) \frac{u_l}{R} P_l (\cos \Theta) \hat{\bm R} + \left  (  \frac {u_l}{R} + \partial_R u_l \right) [\partial_\Theta P_l (\cos \Theta)] \hat{\bm \Theta} \,, {\rm~~where}\\
u_{l} & = \epsilon u^{(1)}_{l} + \epsilon^2 u^{(2)}_{l} + O(\epsilon^3) \,.
\end{align}
We see that the $\zeta_\Theta$ is simply related to $\zeta_R$ through the requirement that ${\bm \zeta}$ be volume-preserving. Comparing Eq.~\eqref{eq:zeta1} to Eq.~\eqref{eq:zetadef}, we immediately find that
\begin{align} 
u^{(1)}_l =  - \omega_0^2 \frac{R^3 }{6 \mathfrak g} \delta_{l2} \,.
\end{align}
Substituting this into Eqs.~\eqref{eq:zeta2} and~\eqref{eq:zetadef} allows us to solve for the functions $u^{(2)}_l$. The angular terms are of the form $(P_2)^2$ and $(\partial_\Theta P_2)^2$, which when re-expanded in terms of Legendre polynomials couple only to $l = 2, 4$. We can pick off each of these terms by integrating Eq.~\eqref{eq:zeta2} with the appropriate Legendre polynomial and weight $d \mu = d[\cos \Theta]$. Orthogonality then gives
\begin{align}
u^{(2)}_2  = - \frac{\omega_0^4}{84}\frac{R^4 (8 - n )}{ \mathfrak g^2} \, {\rm~~and}~~
u^{(2)}_4  = \frac{3\omega_0^4}{350} \frac{ R^4(1- n )}{\mathfrak g^2}\,;
\end{align} 
all the other $u^{(2)}_l$ vanish. We have defined $n = d \ln \mathfrak g / d\ln R$, which proves to be a convenient short hand in the estimates in \S\ref{sec:Compute}. When $\mathfrak g$ has a simple power-law dependence on $R$, $n$ is just its constant power law index. Note that the $l=0$ contribution, which is $\Theta$ independent, represents the correction to the spherically symmetric potential $\Phi_0(R)$. Because of this we also have from the matching that 
\begin{align}
\Phi(R) & = \Phi_0 (R) + \epsilon^2 V (R) + O(\epsilon^3) \,, {\rm~~where}\\
 V (R)& = -\frac{\omega_0^4}{10} \frac{R^3 (6-n ) }{\mathfrak g} \,.
\end{align}
In order to have the generator of the transform $\psi$ we simply need to replace the coordinates $(R,\Theta)$ with $(r,\theta)$, since we have already accounted for the sign change in reversing the flow. As such, we have completed the construction of $\psi$ up to $O(\epsilon^2)$. 

\subsection{Understanding the transform}
\label{sec:Understanding}

\begin{figure}
\centering 
  \fbox{\includegraphics[width=1.6in]{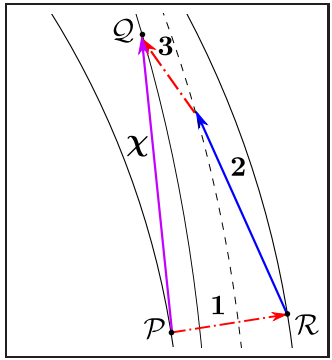}}
   \caption{Depiction of the nonlinear tidal response ${\bm \chi}$ as a sum of virtual displacements. The purely radial displacement ${\bm 1}$ changes the density and pressure of the fluid elements. The virtual displacements ${\bm 2}$ and ${\bm 3}$ are volume-preserving, and correspond to the linear $l=2$ tide and the volume-preserving component of the nonlinear tide, respectively.}
  \label{fig:Transform}
\end{figure}

We have an expression for the volume-preserving transform which takes the tidally perturbed star A into the spherically symmetric star B. We have seen that star B is more weakly perturbed than star A, and this allows us to match orders in perturbation in a way that gives useful relations. First though, it is useful to step back and consider the physical intuition underlying the transformation. Figure~\ref{fig:Transform} illustrates the principles behind the transformation. Consider an element displaced from point $\mathcal P$ to point $\mathcal Q$ by the tidal displacement ${\bm \chi}$. As we noted before, this tidal displacement takes the star to a configuration where elements along contours of equal potential have the same density and pressure. Point ${\mathcal Q}$ lies on one such contour. The displacement field ${\bm \chi}$ can be decomposed as the sum of certain virtual displacement fields, appropriate members of which are marked as ${\bm 1}$, ${\bm 2}$ and ${\bm 3}$ in the figure. These virtual displacements are chosen to have some nice properties. 

First, each virtual displacement is chosen to be adiabatic. This is possible since their sum, ${\bm \chi}$, is adiabatic. Second, we choose ${\bm 2}$ and ${\bm 3}$ such that they both preserve the volume of fluid elements. The pressure and density of an element at $\mathcal P$ is not necessarily equal to the pressure and density of the element once it is displaced to $\mathcal Q$, as the tidal deformation ${\bm \chi}$ is not volume-preserving beyond linear order. With this choice for ${\bm 2}$ and ${\bm 3}$, the entire volume change of an element, and the related pressure and density change, occurs during displacement ${\bm 1}$. Third, the displacement ${\bm 2}$ is chosen to be the linear part of ${\bm \chi}$ (the so-called linear tide). This is possible because the linear tide is a volume-preserving deformation. Given this choice, the displacements ${\bm 1}$ and ${\bm 3}$ make up the nonlinear part of ${\bm \chi}$. Fourth, ${\bm 3}$ is chosen to be that part of the nonlinear tide which preserves volumes, such that the remaining displacement, ${\bm 1}$, takes the initial spherical potential contour that $\mathcal P$ lies on to another spherical contour. Hence, the displacement ${\bm 1}$ is a radial displacement, which takes $\mathcal P$ to a point $\mathcal R$. 

We are now faced with the question of what sort of perturbing fields we would need to apply to make the elements follow these virtual displacements. We have seen that ${\bm 2}$ and ${\bm 3}$ correspond to a pure coordinate transformation (they make up $\psi^{-1}$), so that the potential $\mathcal V$ of the element is frozen in during these displacements. The overall stellar structure naturally flows forward along with the transformation. Hence the potential at $\mathcal R$ is same as the potential at $\mathcal Q$. This differs from the background star's potential at $\mathcal P$, and the difference can be seen as a spherically symmetric perturbing potential causing the elements of the star to deform along ${\bm 1}$.

We could judge this decomposition scheme solely on its power to illuminate the underlying components of the equilibrium tide ${\bm \chi}$, but its real value manifests in the study of small displacements on top of this, away from equilibrium. This is because these weak displacements, say ${\bm \eta}$, which are originally on a star deformed to $O(\epsilon)$, when pulled back along ${\bm 3}$ and ${\bm 2}$ pick up additional corrections of $O(\epsilon)$ and higher while the transformed star is more weakly deformed [perturbed at $O(\epsilon^2)$]. We can explicitly compute these corrections to the displacements from the pullback. In order to get the $\epsilon^2 \eta^2$ part of the energy, we have to look at both of the functionals $f_3$ and $f_4$ in the original picture. Here we find that we only need $f_3$.

\subsection{Potential of star B}
\label{sec:PotentialB}

Now that we have an explicit form for the transformation between star A and star B, and a better understanding of the motivation behind this transform, we can consider the potential $\mathcal V$ of star B. Star B sits in a spherically symmetric gravitational potential, and its fluid elements have been adjusted by the combination of the displacement field ${\bm \chi}$ and the coordinate transform. In total, the star has undergone a radial deformation ${\bm \sigma}$ (before called ${\bm 1}$), sourced by the external potential $\epsilon^2 V$, so that ${\bm x} \to {\bm X} = {\bm x} + {\bm \sigma}$. As before, we have for the new configuration
\begin{align}
\mathcal V({\bm \sigma}) = & \int d^3 X \, \rho ({\bm X}) \left[ \mathcal E_{\rm int} ({\bm X}) + \Phi_0({\bm X}) + \epsilon^2 V({\bm X}) \right] =  \int d^3 x \, \rho ({\bm x}) \left[ \mathcal E_{\rm int} ({\bm x}+{\bm \sigma}) + \Phi_0({\bm x}+{\bm \sigma}) + \epsilon^2 V({\bm x}+{\bm \sigma}) \right] \notag \\
=& \mathcal V(0) + \int d^3x \, \rho({\bm x}) \left[ \frac 12 {\bm \sigma}\cdot {\bm C}\cdot {\bm \sigma} + \frac{1}{3!} f_3({\bm \sigma},{\bm \sigma},{\bm \sigma}) + \hdots + \epsilon^2 V({\bm x}) + \epsilon^2 {\bm \sigma}\cdot {\bm \nabla} V + \epsilon^2 \frac12 {\bm \sigma} \cdot ({\bm \sigma} \cdot {\bm \nabla}) {\bm \nabla} V + \hdots \right]\,.
\end{align}
Expanding ${\bm \sigma}$ in the mode basis of the unperturbed star, ${\bm \sigma} = \sum \sigma_a {\bm \xi}_a$, we can write the potential as
\begin{align}
\mathcal V =& \mathcal V(0) +  \int d^3 x \, \rho \, \epsilon^2 V+ E_0 \sum \left[ \frac 12 \sigma_a^2 - \frac 13 \kappa_{abc} \sigma_a \sigma_b \sigma_c  - \epsilon^2 V_a \sigma_a - \epsilon^2 \frac 12 V_{ab} \sigma_a \sigma_b + \hdots \right] \,,
\end{align}
with
\begin{align}
V_a =  -\frac{1}{E_0} \int d^3 x \, \rho\,  {\bm \xi}^*_a \cdot {\bm \nabla} V \,{\rm~~and~~}
\label{eq:VabMode}
V_{ab} =  -\frac{1}{E_0} \int d^3 x \, \rho\, {\bm \xi}_a \cdot ({\bm \xi}_b \cdot {\bm \nabla}) {\bm \nabla} V \,.
\end{align}
Defining an expansion in $\epsilon$ for $\sigma_a$ which must begin at $O(\epsilon^2)$ because the perturbing tidal field enters only at this order,
\begin{align}
\sigma_a = \epsilon^2 \sigma_a^{(2)} + O (\epsilon^3) \,,
\end{align}
we can again solve for $\sigma_a$ for this static configuration by minimizing the variation in $\mathcal V$ with respect to the amplitude $\sigma_a$. At leading order, we simply have
\begin{align}
\label{eq:sigmaSln}
\sigma_a^{(2)} = V_a \,.
\end{align}
It is worth noting here that since ${\bm \sigma}^{(1)} = 0$, we immediately have that
\begin{align}
\label{eq:chi1zeta1}
{\bm\chi}^{(1)} = - {\bm \zeta}^{(1)} \,.
\end{align}
To see this, we note that by definition the displacement of a point ${\bm x}$ in the unperturbed star by ${\bm \sigma}$ is the same as the composition of a displacement by ${\bm \chi}$ and then the application of the coordinate transform $\psi$. Keeping only the leading order terms in $\epsilon$ this gives
\begin{align}
{\bm x} + {\bm \sigma}({\bm x}) =  {\bm x} + O(\epsilon^2) =   {\bm x} + \epsilon{\bm \chi}^{(1)}({\bm x}) + \epsilon{\bm \zeta}^{(1)}({\bm x}) + O(\epsilon^2) \,,
\end{align}
from which Eq.~\eqref{eq:chi1zeta1} follows.

As before, we now consider further perturbations to star B, which we denote ${\bm \eta}_S$. From the original unperturbed star we have a displacement
\begin{align}
\label{eq:xiSpherical}
{\bm x} \to & {\bm x} + {\bm \xi} = {\bm x} + {\bm \sigma} + {\bm \eta}_S \,{\rm~~where} \\
\label{eq:etaS}
{\bm \eta}_S = & {\bm J}_\psi \cdot {\bm \eta} \,.
\end{align}
Here, ${\bm \eta}$ is the form of the corresponding oscillations of star A, which must be transformed into the coordinates ${\bm X}$ of the spherical star B if we are to consider further displacements beyond $\sigma$. This is the origin of the Jacobian transformation matrix ${\bm J}_\psi$ in Eq.~\eqref{eq:etaS}. In principle, Eq.~(\ref{eq:etaS}) should include terms of order ${\bm\eta}^2$; however, since star B is an equilibrium solution, the potential energy contains no terms of first order in ${\bm\eta}_S$ and the leading dependence is ${\bm\eta}_S^2$. Therefore to compute the potential energy to order ${\bm\eta}^2$, we need only obtain ${\bm\eta}_S$ to linear order in ${\bm \eta}$.

Now we consider the potential $\mathcal V$ of star B when the additional perturbations ${\bm J}_\psi \cdot {\bm \eta}$ are present. We define the expansion of the Jacobian in orders of $\epsilon$, 
\begin{align}
\label{eq:VectorJac}
{\bm J}_\psi = {\bm 1} + \epsilon {\bm J}^{(1)}_\psi + \epsilon^2 {\bm J}^{(2)}_\psi + O (\epsilon^3) \,,
\end{align}
and its expansion in the mode basis ${\bm \xi}_a$ of the original star,
\begin{align}
J_{ab} = \frac{\omega_a^2}{E_0} \int d^3 x \, \rho \,   {\bm \xi}^*_a \cdot {\bm J}_\psi \cdot {\bm \xi}_b \,,
\end{align}
so that the basis coefficients are naturally transformed as
\begin{align}
{\bm \eta}_S = \sum \eta_{S,a} {\bm \xi}_a = \sum J_{ab} \eta_b {\bm \xi}_a = \sum (\delta_{ab} + \epsilon J_{ab}^{(1)} + \epsilon^2 J_{ab}^{(2)}) \eta_b {\bm \xi}_a \,.
\end{align}

Taking all of this into account, we have for star B the mode expansion of the potential
\begin{align}
\mathcal V( {\bm \sigma} + {\bm \eta}_S) = & \mathcal V(0) +  \int d^3 x \rho \epsilon^2 V + \frac{E_0}{2} \sum \left[  \eta_a^2 + \epsilon  \left(J^{(1)}_{ab} + J^{(1)}_{ba}\right)\eta_a \eta_b+ \epsilon^2 \left( J^{(1)}_{ca} J^{(1)}_{cb} + J^{(2)}_{ab} + J^{(2)}_{ba}\right)\eta_a \eta_b  \right. \notag \\
& \left. -  2\kappa_{abc} \eta_a \eta_b \sigma_c  - \epsilon^2  V_{ab} \eta_a \eta_b + \hdots \right] \nonumber \\
=&   \mathcal V(0) +  \int d^3 x \rho \epsilon^2 V + \frac{E_0}{2} \sum \left[ \eta_a^2 + \epsilon \left(J^{(1)}_{ab} + J^{(1)}_{ba}\right)\eta_a \eta_b
\right.\nonumber \\ & \left.
+ \epsilon^2 \left(J^{(1)}_{ca} J^{(1)}_{cb} +J^{(2)}_{ab} + J^{(2)}_{ba} -  2\kappa_{abc}  V_c -V_{ab}\right) \eta_a \eta_b \right],
\label{eq:TransformedV}
\end{align}
where again in the first equality we have already eliminated the terms which vanish when the solution for $\sigma_a$ is inserted and terms higher order in $\epsilon$ or $\eta$ than we are considering. In the second equality we have substituted the solution~\eqref{eq:sigmaSln} for $\sigma_a$ and collected terms according to their order in $\epsilon$.

\subsection{Matching stars A and B in the mode basis}
\label{sec:Matching}

We have computed an expression for $\mathcal V$ in the mode basis using two different methods, and we can now equate these expressions. For star A, $\mathcal V$ is given by the Lagrangian of Eq.~\eqref{eq:ModeL}, through $\mathcal V = \sum \dot \eta_a^2/\omega_a^2- \mathcal L$. Matching this to Eq.~\eqref{eq:TransformedV} order by order in $\epsilon$ and $\eta$, we get the following conditions on the three- and four-mode coupling terms.

At order $\epsilon$, matching terms gives
\begin{align}
\label{eq:FirstMatch}
U_{ab}+ \sum 2 \kappa_{abc} \chi_c^{(1)} = -\left( J^{(1)}_{ab} + J^{(1)}_{ba}\right)\,,
\end{align}
which results in an expression for $\kappa_{abc} \chi_c^{(1)} $ in terms of the Jacobian transform and the tidal potential. Meanwhile, matching the $\epsilon^2$ terms gives
\begin{align}
\label{eq:SecondMatch}
\sum \left( 2 \kappa_{abc} \chi_c^{(2)} + 3 \kappa_{abcd} \chi_c^{(1)} \chi_d^{(1)} \right) = 
- \sum \left(J^{(1)}_{ca} J^{(1)}_{cb} + J^{(2)}_{ab} + J^{(2)}_{ba} -  2\kappa_{abc}  V_c -V_{ab} \right) \,.
\end{align}
In both of Eqs.~\eqref{eq:FirstMatch} and~\eqref{eq:SecondMatch}, the indices $(a,b)$ refer to particular modes and are not summed over.

\subsection{Expressions for the Jacobian in the mode basis}

Before using the matching conditions in Eqs.~\eqref{eq:FirstMatch} and~\eqref{eq:SecondMatch}, we derive the explicit equations for the Jacobian that transforms the perturbations ${\bm \eta}$ on star A to the perturbations ${\bm \eta}_S$. This is achieved by considering the difference between two series of active transformations as depicted in Fig.~\ref{fig:Models}, using a fixed background coordinate system. We begin by considering a fluid element at a point ${\bm x}$ in the unperturbed star. We then consider the application of the tidal perturbation, which takes ${\bm x} \to {\bm x} + {\bm \chi}({\bm x})$, followed by the coordinate transform $\psi$ acting on this element, which acts on the element at its displaced position ${\bm x} + {\bm \chi}({\bm x})$. This takes us across the top row of Fig.~\ref{fig:Models}, and acts to transform the initial point as
\begin{align}
\label{eq:Path1}
{\bm x} \to {\bm x} + {\bm \chi}({\bm x}) + \left. {\bm\zeta} \right|_{ {\bm x} + {\bm \chi}({\bm x})} + \frac12 \left. ({\bm \zeta}\cdot {\bm \nabla}){\bm \zeta}\right|_{ {\bm x} + {\bm \chi}({\bm x})} \,.
\end{align}

Next, consider the series of transforms which first takes ${\bm x}$ to the corresponding point in star A, ${\bm x} \to {\bm x} + {\bm \chi}({\bm x}) +  {\bm \eta}({\bm x})$. Since we wish to express all perturbations in Lagrangian coordinates, the ${\bm \eta}$ is written as a function of the position of the original, unperturbed element. We follow this pair of displacements by the coordinate transform $\psi$, which acts on the actual position of the element. This gives us the position of the element in star B,
\begin{align}
\label{eq:Path2}
{\bm x} \to {\bm x} + {\bm \chi}({\bm x})  + {\bm \eta}({\bm x}) + \left. {\bm\zeta} \right|_{ {\bm x} + {\bm \chi}({\bm x})+ {\bm \eta}({\bm x})} + \frac12 \left. ({\bm \zeta}\cdot {\bm \nabla}){\bm \zeta}\right|_{ {\bm x} + {\bm \chi}({\bm x})+ {\bm \eta}({\bm x})} \,.
\end{align}
The difference between Eqs.~\eqref{eq:Path2} and~\eqref{eq:Path1} is simply ${\bm \eta}_S({\bm x}) = {\bm J}_\psi {\bm \eta}({\bm x})$. Taking this difference, and expanding our expressions in terms of small ${\bm \eta}$, we have
\begin{align}
{\bm J}_\psi {\bm \eta} =& {\bm \eta}({\bm x}) + \left. {\bm \eta}({\bm x}) \cdot ({\bm \nabla} {\bm \zeta})\right|_{ {\bm x} + {\bm \chi}({\bm x})} + \frac 12  \left. {\bm \eta}({\bm x}) \cdot [ {\bm \nabla} ({\bm \zeta}\cdot {\bm \nabla}){\bm \zeta}\right]_{ {\bm x} + {\bm \chi}({\bm x})} + O(\eta^2)\,.
\end{align}
Further expanding out ${\bm \chi}$ and ${\bm \zeta}$ in terms of $\epsilon$, this expression becomes
\begin{align}
\label{eq:VecJacRaw}
{\bm J}_\psi {\bm \eta} =& {\bm \eta}+ \epsilon ({\bm \eta} \cdot {\bm \nabla}) {\bm \zeta}^{(1)}+ \epsilon^2 \left( ({\bm \eta} \cdot {\bm \nabla}) {\bm \zeta}^{(2)}+ {\bm \eta} \cdot [({\bm \chi}^{(1)} \cdot{\bm \nabla}){\bm \nabla} {\bm \zeta}^{(1)})] + \frac 12 ({\bm \eta} \cdot {\bm \nabla}) ({\bm \zeta}^{(1)}\cdot {\bm \nabla}){\bm \zeta}^{(1)} \right) + O (\eta^3)\,.
\end{align}
Here, all terms are evaluated at the base point ${\bm x}$. This allows us to simply read off the Jacobian of Eq~\eqref{eq:VectorJac}. It is useful to recall that ${\bm \chi}^{(1)} = - {\bm \zeta}^{(1)}$, which allows us to simplify the Jacobian somewhat:
\begin{align}
{\bm J}^{(1)}_\psi {\bm \eta}  =  ({\bm \eta} \cdot {\bm \nabla}) {\bm \zeta}^{(1)} \,{\rm~~and~~} 
{\bm J}^{(2)}_\psi {\bm \eta}  = ({\bm \eta} \cdot {\bm \nabla}) {\bm \zeta}^{(2)} + [({\bm\eta} \cdot {\bm \nabla}) {\bm \zeta}^{(1)}]\cdot {\bm \nabla}  {\bm \zeta}^{(1)} - \frac 12  ({\bm \eta} \cdot {\bm \nabla}) ({\bm \zeta}^{(1)}\cdot {\bm \nabla}){\bm \zeta}^{(1)} \,.
\end{align}
To simplify Eq.~\eqref{eq:VecJacRaw} we have temporarily resorted to a Cartesian basis in order to commute the covariant derivatives.

Now that we have an explicit expression for the Jacobian, we can express it in the mode basis using the expansion for ${\bm \eta}$. The result, for the first and second order terms, is
\begin{align}
\label{eq:ModeJac1}
 J_{ab}^{(1)}&=\frac{\omega_a^2}{E_0}\int d^3x \rho\, {\bm \xi}_a\cdot({\bm \xi}_b\cdot{\bm \nabla}){\bm \zeta}^{(1)} \,{\rm~~and} \\
\label{eq:ModeJac2}
J_{ab}^{(2)}&=\frac{\omega_a^2}{E_0}\int d^3x \rho \, {\bm \xi}_a\cdot({\bm \xi}_b\cdot{\bm \nabla})\left[{\bm \zeta}^{(2)}-\frac{1}{2}({\bm \zeta}^{(1)}\cdot { \bm \nabla}){\bm \zeta}^{(1)}\right]  + {\bm \xi}_a\cdot\left( \left[ ({\bm \xi}_b\cdot{\bm \nabla}){\bm \zeta}^{(1)} \right] \cdot {\bm \nabla}{\bm \zeta}^{(1)} \right)\,.
\end{align}
For our initial check of the mode stability in \S\ref{sec:ModeStability}, we note here that for a particular high-order mode, $\xi_a \sim \omega_a^{-1}$, and so we have the useful fact that for a particular pair of modes $(a,b)$,
\begin{align}
J^{(i)}_{ab} \sim \frac{\omega_a}{\omega_b}
\end{align}
so long as the angular integrations satisfy selection rules and the contraction of indices in Eqs.~(\ref{eq:ModeJac1},\ref{eq:ModeJac2}) do not lead to a much smaller value. This fact is made more explicit in \S\ref{sec:Compute}.

\subsection{Lagrangian perturbations of star A revisted}
\label{sec:ModeStabilityRevisited}

With our matching results from \S\ref{sec:Matching}, we can return to the expressions for the perturbed eigenvalues from \S\ref{sec:ModeStability}. We focus on the case of interest for the instability proposed by WAB, that of a high-order $(p,g)$ mode pair with comparable wave numbers and widely spaced frequencies, $\omega_p \gg \omega_g$. Working in the rescaled mode amplitudes $\eta_a' = \eta_a/\omega_a$, consider the eigenvalues of the matrix $\mathcal M$ in the Lagrangian~\eqref{eq:MatrixL}. Substituting the matching conditions~\eqref{eq:FirstMatch} and~\eqref{eq:SecondMatch} into $\mathcal M$ as given by Eqs.~\eqref{eq:masspp}--\eqref{eq:massgg}, we have now that
\begin{align}
\mathcal M_{pp} = & \omega_p^2 + 2 \epsilon \omega_p^2 J^{(1)}_{pp}  + \epsilon^2 \omega_p^2 \left[ \sum\left( J^{(1)}_{ap} J^{(1)}_{ap}  -  2\kappa_{ppa}  V_a \right) + 2 J^{(2)}_{pp} -V_{pp} \right] \,, 
\label{eq:Mpp}\\
\mathcal M_{pg} = & \mathcal M_{gp} = \epsilon \,  \omega_p \omega_g \left( J^{(1)}_{pg} + J^{(1)}_{gp} \right) \,, {\rm~~and} 
\label{eq:Mpg} \\
\mathcal M_{gg} = & \omega_g^2 + 2 \epsilon \omega_g^2 J^{(1)}_{gg}  + \epsilon^2 \omega_g^2 \left[ \sum\left( J^{(1)}_{ag} J^{(1)}_{ag}  -  2\kappa_{gga}  V_a \right) + 2 J^{(2)}_{gg} -V_{gg} \right] \,.
\end{align} 
The resulting perturbed eigenvalues are similarly re-expressed in terms of the Jacobian and the potential $V$. Let us focus on the smaller frequency, which is perturbed to a lower (and possibly negative value):
\begin{align}
\label{eq:omegaM}
\frac{\omega_-^2}{\omega_g^2} = &1 + 2 \epsilon J_{gg}^{(1)} + \epsilon^2 \left[\left(J^{(1)}_{pg}\right)^2 +\left(J^{(1)}_{gg}\right)^2 + 2 J^{(2)}_{gg} -V_{gg} -\sum 2\kappa_{gga}  V_a \right] - \epsilon^2 \frac{\omega_p^2}{\omega_p^2- \omega_g^2} \left( J^{(1)}_{pg} + J^{(1)}_{gp} \right)^2\,.
\end{align}
Keeping in mind the origin of each of the terms, and our estimate that $J^{(i)}_{ab} \sim \omega_a/\omega_b$, we can see how Eq.~\eqref{eq:omegaM} encodes a potential instability, and the particular manner in which it is actually canceled. The final $O(\epsilon^2)$ term in Eq.~\eqref{eq:omegaM} is just the three-mode term $\kappa_{pga} \chi_a^{(1)}$ in Eq.~\eqref{eq:omegaminus}, expressed in the language of Jacobians. When the frequency of the $p$-mode is much larger than that of the $g$-mode, the size of this term is $\sim (J^{(1)}_{pg})^2 \sim (\omega_p^2/\omega_g^2) \gg 1$. In principle, this can overcome the $\epsilon^2$ suppression when the tidal field is relatively strong (but with $\epsilon \ll 1$), and overwhelm the order-unity contribution from the restoring force. However, when we express the perturbation to the diagonal terms [entering from the four-mode and $\kappa_{abc}\chi^{(2)}_c$ terms in Eq.~\eqref{eq:omegaminus}] in terms of Jacobians, we observe that it contains an identical contribution with the opposite sign. 

Before we explicitly write down the remaining terms, it is worthwhile to pause and consider the impact of our coordinate transformation. The three- and four- mode terms in Eq.~\eqref{eq:omegaminus} both affect the eigenvalues equally. Independently calculating their individual contributions would have involved careful book-keeping in order to accurately track the cancellation of large terms. Instead, our approach has confirmed the intuition developed from the toy model of Sec.~\ref{ss:toy}, and illuminated the fact that large coupling terms can arise from rotations of modes into each other.

We now write down the terms that remain after the large cancellations. Expanding the prefactor of the final term in Eq.~\eqref{eq:omegaM},
\begin{align}
\frac{\omega_p^2}{\omega_p^2- \omega_g^2}  = \left( 1 + \frac{\omega_g^2}{\omega_p^2} \right) + {\cal O}\left(\frac{\omega_g^4}{\omega_p^4}\right)\,,
\end{align}
we can simplify Eq.~\eqref{eq:omegaM} to
\begin{align}
\label{eq:omegaM2}
\frac{\omega_-^2}{\omega_g^2} = &1 + 2 \epsilon J^{(1)}_{gg} + \epsilon^2 \left[\frac{\omega_g^2}{\omega_p^2}\left(J^{(1)}_{pg}\right)^2+\left(J^{(1)}_{gg}\right)^2 + 2 J^{(2)}_{gg} -V_{gg} -2J^{(1)}_{pg}J^{(1)}_{gp} -\sum 2\kappa_{gga}  V_a  \right] \,.
\end{align}
Our estimate for the size of the Jacobian terms shows us that all the terms are ${\cal O}(1)$ in terms of frequency, except perhaps the $V_{gg}$ and $\kappa_{gga}V_a$ terms. We need to check that these terms are not large for the case of the high-order $(p,g)$-mode coupling. We show this formally and estimate the size of these terms in \S\ref{sec:Compute}.

In both this section and in \S\ref{sec:ModeStability}, we have considered the coupling of a single pair of modes $(p,g)$, but in general $\mathcal M$ contains entries for all modes and their mode-mode coupling terms. Equations~\eqref{eq:Mpp}--\eqref{eq:omegaM2} extend naturally to this case. For example, if we consider a single $g$-mode, there is an off-diagonal matrix entry $\mathcal M_{a g}$ for every other mode $\eta_a$, and this entry couples the $g$-mode to $\eta_a$. In addition, the sum $\sum  J^{(1)}_{ag} J^{(1)}_{ag}$ in $\mathcal M_{gg}$ contains a contribution from each of these other modes. When computing the perturbation to the $g$-mode frequency, the off-diagonal terms all enter as a sum of squared terms analogous to the last term in Eq.~\eqref{eq:omegaM}. In those cases where a mode strongly couples to the $g$-mode, the same leading-order cancellation of large terms occurs mode-by-mode. This means that even in the case where many $p$-modes strongly couple to a single $g$-mode, cancellation between the large parts of the three-mode couplings (squared) and four-mode couplings prevents the $g$-mode from developing an instability.

It is remarkable that up until now, our investigation of the generation of the daughter modes from the tidal perturbation has been formal and general in nature, with no reference made to a particular stellar model. In our estimates we have only needed to note that ${\bm \xi}_a \sim \omega_a^{-1}$ given our normalization of the mode functions.

\section{Estimating the remaining terms}
\label{sec:Compute}

So far, we have shown that the major potential contributions to the instability discussed in WAB cancel out in the limit $\omega_p \gg \omega_g$ for the $(p, g)$ pair of daughter modes. This result alone does not guarantee the stability of the star to the production of the $(p,g)$ mode pair, since there are other terms whose sizes need to be estimated for the modes of interest. We carry out these estimates in this section, relying on the WKB approximation for the mode functions when we need to explicitly compute the size of the various terms in Eq.~\eqref{eq:omegaM}. The WKB approximation is appropriate in this case, since the proposed instability occurs when high order $p$- and $g$-modes have resonant spatial eigenfunctions, a condition which requires large wave numbers $k_p$ and $k_g$ for the modes.

First, though, we develop some confidence in the matching results of \S\ref{sec:Matching} by showing that the three-mode coupling term $\kappa_{pgc} \chi_c^{(1)}$ as derived in WAB can be recovered from our matching equations. We then turn to estimating the size of the remaining terms.

\subsection{The three-mode coupling}

We begin by considering the amplitude $U_a = \chi^{(1)}_a = - \zeta^{(1)}_a$. It is conveniently obtained by considering the leading order part of ${\bm \zeta}$,
\begin{align}
\label{eq:LinearTide}
{\bm \zeta}^{(1)}  = - {\bm \chi}^{(1)} = - \omega_0^2\frac{r^2}{\mathfrak g} \left[ P_2(\cos \theta) \hat{\bm r} + \frac {4-n}{6} \partial_\theta P_2(\cos \theta) \hat{\bm \theta} \right] \,.
\end{align}
When we recall the definition that $\epsilon = Gm/(a^3\omega_0^2)$ this recovers the linear tidal response,
\begin{align}
\epsilon {\bm \chi}^{(1)} = \frac{Gm}{a^3} \frac{r^2}{\mathfrak g} \left[ P_2(\cos \theta) \hat{\bm r} + \frac {4-n}{6} \partial_\theta P_2(\cos \theta) \hat{\bm \theta} \right] \,,
\end{align}
which can be compared to Eqs~(A12) and~(A13) of \citet{Weinberg2012}. Recall that our eigenmodes ${\bm \xi}_a$ have the form in Eq.~\eqref{eq:chiexpansion}, and that here we have specialized to the $m=0$ case. We see that ${\bm \chi}^{(1)}$ has the correct form for an $l=2$ displacement.

Before continuing, it is useful to note that we have frequent need of integrals of the form
\begin{align}
\label{eq:3ModeInt}
I_{abc} = & \int d^3 x \, \rho \, {\bm \xi}_a \cdot ({\bm \xi}_b \cdot {\bm \nabla}) {\bm \xi_c} \,.
\end{align}
With our convention that ${\bm \xi}_a$ take the form in Eq.~\eqref{eq:chiexpansion}, the integral~\eqref{eq:3ModeInt} resolves to
\begin{align}
\label{eq:3ModeIntExp}
I_{abc} = \int dr r^2 \rho(r) &\left[ T_{abc} a_r b_r \frac{d c_r}{d r} + F_{a,bc} \frac{a_r b_h}{r} (c_r - c_h) + F_{b,ac} a_h b_r \frac{d c_h}{d r}  + \frac{a_h b_h}{r} ( G_{c,ab} c_h + F_{c,ab} c_r ) \right].
\end{align}
The angular integrals $T_{abc},\, F_{a,bc},$ and $G_{a,bc}$ are those originally defined in \citet{Wu2001} and match those listed in Eqs.~(A20--A22) of \citet{Weinberg2012}. We give them in Appendix~\ref{sec:CalcDetails}, and they are determined by the angular indices $(l_a,l_b,l_c)$ of the mode vectors in the integral. These integrals vanish unless the angular momentum indices obey the triangle inequality $|l_b - l_c| \leq l_a \leq l_b + l_c$, and in addition the sum $l_a + l_b + l_c$ must be even.

Now we are ready to consider the three-mode coupling term for the case of a pair of $(p,g)$ modes excited by the tide. From Eq.~\eqref{eq:FirstMatch}, we have
\begin{align}
  \sum \kappa_{pgc} \chi_c^{(1)}  = - \frac 12 U_{pg} - \frac 12 \left( J^{(1)}_{pg} + J^{(1)}_{gp}\right)\,.
\end{align}
The coupling $\kappa_{pgc}$ can be large in the case $\omega_p \gg \omega_g$, and when the spatial resonance condition $k_p \simeq k_g$ is met \citep{Weinberg2013}. This means that the $(p,g)$ modes must be high order modes with large wave numbers, and as such we use the WKB approximation \citep{Unno1989,Weinberg2012} to get analytic forms for the eigenfunctions ${\bm \xi}_p$ and ${\bm \xi}_g$. In this case, the radial functions $(p_r, p_h)$ and $(g_r,g_h)$ are
\begin{align}
\label{eq:pmodewkb}
(p_r, p_h)&\simeq \frac{A_p}{\omega_p}\left(\cos{k_p r},\frac{c_s \sin{k_p r}}{\omega_p r}\right)\,{\rm~~and} \\
\label{eq:gmodewkb}
(g_r, g_h)&\simeq \frac{A_g}{\omega_g}\left(\frac{\omega_g \sin{k_g r}}{N},\frac{\cos{k_g r}}{\Lambda_g}\right) \,, 
\end{align}
where $c_s$ is the adiabatic sound speed, $N$ is the Br\"unt-V\"ais\"al\"a frequency,
\begin{align}
N^2&=-\left(\frac{1}{\rho}\frac{d\rho}{dr}-\frac{1}{\Gamma_1 P}\frac{dP}{dr}\right)\mathfrak g,
\end{align}
$\Gamma_1$ is the adiabatic index, $\Lambda_a^2=l_a(l_a+1)$, and we set
\begin{align}
    (k_p,k_g) &\simeq \left(\frac{\omega_p}{c_s},\frac{\Lambda_g N}{r\omega_g}\right), &
    A_{p,g}&=\sqrt{\frac{E_0\alpha_{p,g}}{\rho r^2}}\,,   &  \alpha_p&=\left(c_s\int c_s^{-1} dr\right)^{-1}\,, &{\rm and}& &
    \alpha_g&=\frac{N}{r}\left(\int N d\ln{r}\right)^{-1}\,.
\end{align}
In the WKB approximation, $k_p \simeq k_g$ implies that  $\omega_p \omega_g \simeq \Lambda_g N c_s/ r$. The key consideration for estimating the largest terms is that for the $p$-mode, $p_r$ is the dominant component of the mode (the acoustic modes are mostly radial) and for the $g$-mode, $g_h$ is the dominant component (the gravity wave modes are mostly horizontal). 

When we have need for a specific stellar model, we use rough approximations to the neutron star model \citep{2009PhRvL.103r1101S} used by WAB, generated with the Skyrme-Lyon equation of state \citep[SLy4;][]{1998NuPhA.635..231C}. In this model the density is approximately constant throughout the core of the neutron star, where the coupling $\kappa_{pgc}$ is large. When this is true the gravitational acceleration grows linearly with the radius, $\mathfrak g \simeq 4 \pi G \rho r/3$.  In addition, $c_s$ is roughly constant throughout the core of the star and so $c_s \simeq \omega_0 R_*$. We also use the expression for $N$ from \citet{Reisenegger1992}, derived to leading order in small electron fraction $Y_e$,
\begin{align}
N \simeq \frac{\mathfrak g}{c_s}\sqrt{\frac{Y_e}{2}} \,.
\end{align}
This leads us to see that $N c_s /r$ is constant, and moreover that $N \sim \omega_0 r/R_*$, since $\omega_0^2 \simeq G \rho$ for a nearly constant density. Taken together with the condition that the wave numbers for the $(p,g)$ modes are nearly equal, we have that $\omega_p \omega_g \sim \omega_0^2 $. In this model, Eq.~\eqref{eq:LinearTide} shows that $\chi^{(1)}_{r}\sim \chi^{(1)}_{h} \sim r$. These approximations are expected to hold for a variety of neutron star models (see WAB \S3.3).

We begin our computation of the three-mode coupling constant by evaluating $U_{pg}$, 
\begin{align}
U_{pg} =&  - \frac{\omega_0^2}{E_0} \int d^3 x \, \rho \, {\bm \xi}_p \cdot ({\bm \xi}_g \cdot{\bm \nabla }){\bm \nabla } [r^2 P_2(\cos \theta) ]  \notag \\
=&  - \frac{\omega_0^2}{E_0} \sqrt{\frac{4\pi}{5}} \int dr \, r^2 \rho \left[ 2 p_r g_r T_{pg2} + p_r g_h F_{p,g2} + p_h g_r F_{g,p2} + p_h g_h (2 F_{2,pg}+ G_{2,pg})\right] \notag
\\ \simeq &  - \frac{\omega_0^2}{E_0} \sqrt{\frac{4\pi}{5}}  \int dr r^2 \rho \, p_r g_h F_{p,g2} \sim \frac{\omega_0^2}{\omega_p \omega_g} \sim {\cal O}(1) \,,
\end{align}
where in the third line terms have been dropped since they are higher order in $\omega_g/\omega_0$ or $\omega_0/\omega_p$. We see that $U_{pg}$ contains no large terms. 

Similarly, we can consider the leading order Jacobian terms, using Eqs.~\eqref{eq:ModeJac1} and~\eqref{eq:3ModeIntExp} to write them as,
\begin{align}
\label{eq:JacI}
J_{ab}^{(1)} = & \frac{\omega_a^2}{E_0} I_{a b \zeta} = -\frac{\omega_a^2}{E_0} I_{a b \chi^{(1)}} \,,
\end{align}
where the subscripts $\zeta$ and $\chi^{(1)}$ indicate the substitution of the corresponding displacement for the mode function ${\bm \xi}_c$ in the integral $I_{abc}$. The radial functions $\chi_r$ and $\chi_h$ are defined as in Eq.~\eqref{eq:chiexpansion}, with $l = 2$ and $m=0$, so that the largest contributions to the Jacobian terms are
\begin{align}
J_{pg}^{(1)} = & - \frac{\omega_p^2}{E_0} \int dr \, r^2 \rho \,  F_{p,g2} \frac{p_r g_h}{r} (\chi_r^{(1)} - \chi_h^{(1)}) \sim \frac{\omega_p}{\omega_g} \,
{\rm~~and}
\\
J_{gp}^{(1)} = & - \frac{\omega_g^2}{E_0} \int dr \, r^2 \rho \,  F_{g,p2} \, g_h p_r \frac{d\chi_h^{(1)}}{dr} \sim \frac{\omega_g}{\omega_p} \,.
\end{align}
At leading order, then,
\begin{align}
  \sum \kappa_{pgc} \chi_c^{(1)} \sim   \frac{\omega_p^2}{2E_0} \int dr \, r^2 \rho \,  F_{p,g2} \frac{p_r g_h}{r} (\chi_r^{(1)} - \chi_h^{(1)}) \sim \frac{\omega_p}{\omega_g} \,.
\end{align}
This matches the leading order terms in WAB, and arises solely from our consideration of the volume-preserving transform. As we have shown, a correction involving the four-mode coupling term cancels its influence on the eigenfrequencies. We now show that none of the remaining terms are large enough to produce a potential instability.

\subsection{Size of the remaining terms}
\label{sec:RemTerms}

We have calculated the correction to the frequency of the almost-neutrally stable $g$-modes in the presence of the tidal deformation, in Eq.~\eqref{eq:omegaM2}. We can write the relative corrections as
\begin{align}
  \frac{\omega_-^2 - \omega_g^2}{\omega_g^2} = & 2 \epsilon J^{(1)}_{gg} + \epsilon^2 \left[ \left\{ \frac{\omega_g^2}{\omega_p^2}\left(J^{(1)}_{pg}\right)^2+\left(J^{(1)}_{gg}\right)^2 -2J^{(1)}_{pg}J^{(1)}_{gp} + 2 J^{(2)}_{gg}\right\} - \left\{ V_{gg} + \sum 2\kappa_{gga}  V_a  \right\} \right] \,.
\label{eq:deltaomega2}
\end{align}
The corrections can be divided into two classes - a set of Jacobian terms, and a set of potential terms. 

We have previously encountered Jacobian terms while estimating the three-mode coupling. For high-order modes $a$ and $b$, the leading order dependence of the first-order Jacobian $J^{(1)}_{ab}$ on the frequencies is $J^{(1)}_{a b} \sim {\cal O}(\omega_a/\omega_b)$. From this observation, we see that all the terms involving a first-order Jacobian in Eq.~\eqref{eq:deltaomega2} are ${\cal O}(1)$ in the large frequency ratio $\omega_p/\omega_g$.

The second-order Jacobian term $J^{(2)}_{ab}$ is given by Eq.~\eqref{eq:ModeJac2}. We observe that it has the same frequency dependence as the first-order Jacobian, with a prefactor of $\omega_a^2/E_0$ and the two eigenfunctions ${\bm \xi}_a$ and ${\bm \xi}_b$ present in the integrand. As we did for the first order Jacobian, we recall that for high-order modes the WKB eigenfunctions have a size $\xi_a \sim \sqrt{E_0}/ \omega_a$, as seen in Eqs.~\eqref{eq:pmodewkb} and~\eqref{eq:gmodewkb}. Hence, the second-order Jacobian term is ${\cal O}(1)$ in the frequency ratio $\omega_p/\omega_g$. 

Having established that all the Jacobian terms in Eq.~\eqref{eq:omegaM2} are of the form $\epsilon$  or $\epsilon^2$ times terms of ${\cal O}(1)$ in the frequency ratio, we turn to the potential terms in Eq.~\eqref{eq:deltaomega2}. Using Eq.~\eqref{eq:sigmaSln}, we can write
\begin{align}
-\epsilon^2 \left(V_{gg} + \sum 2\kappa_{gga}V_a \right)  = -\epsilon^2 \left(V_{gg} + \sum 2\kappa_{gga}\sigma^{(2)}_a \right) 
= -\epsilon^2 \left(V_{gg} + 2\kappa_{gg\sigma} \right)\,,
\end{align}
where in the last equality we have used the notation introduced in Eq.~\eqref{eq:JacI}, and further dropped the superscript on $\sigma$ with the understanding that we are using the leading $\epsilon^2$ term in the expansion of the displacement. The key point is that $\epsilon^2 \kappa_{gg\sigma}$ can be computed by substituting the displacement ${\bm \sigma}$ in place of the eigenfunction ${\bm \xi}_c$ in the definition~\eqref{eq:kappa3} of the three-mode coupling, as $\kappa_{abc}$ is linear in its arguments. 

The nonlinear tidal term $V_{gg}$ is given by Eq.~\eqref{eq:VabMode}
\begin{align}
  V_{gg} = & -\frac{1}{E_0} \int d^3 x \, \rho\, {\bm \xi}_g \cdot ({\bm \xi}_g \cdot {\bm \nabla}) {\bm \nabla} V \nonumber\\
  = & -\sqrt{4\pi} \frac{1}{E_0} \int dr\, r^2 \rho(r) \left( g_r g_r \frac{d^2 V}{d r^2} T_{\sigma gg} + \frac{g_h g_h}{r} \frac{dV}{dr} F_{\sigma, gg} \right) \nonumber\\
  \label{eq:Vgg}
    = & - \frac{1}{E_0} \int dr\, r^2 \rho(r) \left( g_r g_r \frac{d^2 V}{d r^2} + \Lambda_g^2 \frac{g_h g_h}{r} \frac{dV}{dr} \right) \,.
\end{align}
In this expression, we have used expressions for the angular integrals from~\citet{Wu2001}, together with the fact that the radial displacement ${\bm \sigma}$ has the angular quantum numbers $l=0,\,m=0$. The integrals are elaborated upon in Appendix~\ref{sec:CalcDetails}. Note that in this expression for $V_{gg}$, we have a potentially large part $\sim g_h g_h$ due to the large size of the horizontal displacement $g_h$ of the $g$-mode. We see that this term cancels exactly with a part of the three-mode term.\footnote{The expression $V_{gg}$ as defined by Eq.~\eqref{eq:VabMode} is the spherical ($l=0$) counterpart of the nonlinear driving terms $J_{ablm}$ of \citet{Weinberg2012}, as given by their Eq.~(A23). We can anticipate that there are large cancellations with the inhomogeneous part of the three-mode coupling terms, just as it happens for the functions $J_{ablm}$ in \citet{Weinberg2012}.}

The three-mode term $\kappa_{gg \sigma}$ has a covariant form \citep[e.g.][Eq.~4.20]{Schenk2002}. It can be evaluated in terms of the radial and angular parts of the mode eigenfunctions \citep{Wu2001,Weinberg2012}. Before continuing, we should note a complication which arises in the case we are dealing with, namely that of a radial displacement coupled to non-radial displacements. 

The process of expanding and simplifying the expression for the three-mode coupling involves using the equations of motion for the constituent displacements. The equation for the divergence of a displacement of the form of Eq.~\eqref{eq:chiexpansion} is
\begin{align}
 ( {\bm \nabla} \cdot {\bm \xi}_a )_r = & \frac{d a_r}{d r} + \frac{2}{r}a_r - \frac{\Lambda_a^2}{r} a_h \,, \label{eq:ModeDiv} 
\end{align}
where we have written the divergence as $({\bm \nabla} \cdot {\bm \xi}_a)_r$ because we have omitted its spherical harmonic angular dependence, which is absorbed in the angular integrals. The equations of motion for a mode with angular index $l>0$ are
\begin{align}
  \Gamma_1 P ( {\bm \nabla} \cdot {\bm \xi}_a )_r = &  - (\delta P)_r = \rho \mathfrak{g} a_r - \omega_a^2 \rho r a_h + \rho \Phi^\prime  \,{\rm~~and}
\label{eq:ModeAng} \\
\frac{d}{dr} \left[\Gamma_1 P ( {\bm \nabla} \cdot {\bm \xi}_a )_r \right] = & \frac{\Lambda_a^2}{r}\rho \mathfrak{g} a_h - \left(\omega_a^2 + \frac{2\mathfrak{g}}{r} - \frac{d \mathfrak{g}}{d r} \right) \rho a_r + \rho \frac{d \Phi^\prime}{d r} \,.
\label{eq:ModeRad}
\end{align}
The Eulerian perturbation of the potential is denoted by $\Phi^\prime$. Within the Cowling approximation, it has a contribution only from the external driving potential $V$. Spherical symmetry demands that it only has a contribution from the part of the driving potential with the same spherical harmonic dependence as the displacement. For the $g$-mode, the external driving potential $V_g$ vanishes, because the mode's eigenfunction is nonradial, and the potential is spherically symmetric. 

We cannot use Eqs.~(\ref{eq:ModeDiv}--\ref{eq:ModeRad}) for the radial displacement ${\bm \sigma}$ just by setting the angular displacement $\sigma_h$ to zero. Deriving Eq.~\eqref{eq:ModeAng} involves using the angular parts of the equations of motion, which do not exist for a radial displacement. We could still have used Eq.~\eqref{eq:ModeAng} for the radial displacement had we been operating within the hydrostatic approximation, with the Eulerian pressure perturbation $P^\prime = \rho g \sigma_r = -\rho \Phi^\prime$ and the Lagrangian pressure perturbation $\delta P = 0$. We cannot use the approximation, as it is not consistent with our construction of the radial displacement as that part of the tidal displacement ${\bm \chi}$ which changes the volume of the elements. The root of the difference is that radial and angular modes have different analytic structures. For instance, the Lagrangian and Eulerian pressure perturbations $\delta P$ and $P^\prime$ do not necessarily vanish at the center of the star for a radial displacement. This fact is noted in \citet[\S17.6]{Cox}. Another point of view is that for just a single degree of freedom, Eqs.~\eqref{eq:ModeDiv}--\eqref{eq:ModeRad} are overdetermined, and would need to satisfy a consistency condition.

We {\em can} still use Eq.~\eqref{eq:ModeDiv} (the definition of divergence) and Eq.~\eqref{eq:ModeRad} (the radial equation of motion) for the radial displacement $\sigma$, with the Eulerian perturbation to the potential $\Phi^\prime$ given by the external potential $V$ on star $B$. Equation~\eqref{eq:ModeRad} still holds even though $\sigma$ does not represent a normal mode of the star because it represents a force balance: star B is in hydrostatic equilibrium in the modified potential and hence we may use Eq.~\eqref{eq:ModeRad} with $\omega_\sigma=0$.

Keeping this in mind, the starting point for simplifying the three-mode term $\kappa_{gg \sigma}$ is Eqs. (A27)--(A30) of \citet{Weinberg2012}. Our paths diverge from the point where the equations of motion are used. The end goal of our simplification is to get an expression which consists of terms which have dominant contributions of the form $g_r g_r$ and $g_h g_h$, since given the WKB forms of the mode function, these are the forms which pick up a growing contribution as we integrate through the star. In order to do so, we repeatedly use the equations of motion for the nonradial mode ${\bm g}$ and the radial displacement ${\bm \sigma}$, and integrate by parts wherever needed.

The final expression for the combination of coupling terms needed is 
\begin{align}
  \epsilon^2 \left(V_{g g} + 2 \kappa_{\sigma g g} \right) = & \frac{1}{E_0} \int dr \left[ 
  r^2 P \left\{ \Gamma_1 (\Gamma_1 + 1) + \left(\frac{\partial \Gamma_1}{\partial \ln \rho} \right)_s \right\} ({\bm \nabla} \cdot {\bm \sigma}) ({\bm \nabla} \cdot {\bm g})_r ({\bm \nabla} \cdot {\bm g})_r - 4 r \sigma_r \Gamma_1 P ({\bm \nabla} \cdot {\bm g})_r ({\bm \nabla} \cdot {\bm g})_r
  \nonumber \right. \\
    & \hspace{10pt} \left. 
    - 2 r^2 \left( \Lambda_g^2 \omega_g^2 \rho r g_h g_h + 2 g_r \Gamma_1 P ({\bm \nabla} \cdot {\bm g})_r \right) \frac{d}{d r} \left( \frac{\sigma_r}{r} \right) - \rho \mathfrak{g} r^3 g_r g_r \frac{d^2}{d r^2} \left( \frac{\sigma_r}{r} \right)
    \right. \nonumber \\
    & \hspace{10pt} \left. 
    + \left( - \rho \mathfrak{g}^2 r \frac{d}{d r} \left( \frac{\sigma_r}{ \mathfrak{g} } \right) + \rho r \epsilon^2 \frac{d V}{d r}  \right) \left( 2 r g_r ({\bm \nabla} \cdot {\bm g})_r  +  g_r g_r \frac{d \ln{\rho}}{d \ln{r}} \right)
     \right] \,. \label{eq:3ModeRad}
\end{align}
The process of simplifying the terms from their canonical forms, and the cancellation of the large term in the inhomogeneous driving $V_{gg}$ with the three-mode coupling are demonstrated in more detail in Appendix \ref{sec:CalcDetails}.

In estimating the size of the remaining terms in the above expression, we find it useful to approximate the divergence of the $g$-mode by using Eq.~\eqref{eq:ModeAng} in the following way. First we note that $\omega_g^2 g_h$ is small, and that
\begin{align}
({\bm \nabla} \cdot {\bm g})_r \simeq \frac{\rho}{\Gamma_1 P} \mathfrak g \, g_r  =  \frac{\mathfrak g}{c_s^2} g_r \,.
\end{align}
The radial eigenfunction for the $g$-mode is given by Eq.~\eqref{eq:gmodewkb} within the WKB approximation, which involves the Br\"{u}nt-V\"{a}is\"{a}l\"{a} frequency $N$. Recall that for typical equations of state $N$ and the acceleration due to gravity $\mathfrak{g}$ grow nearly linearly with radius till well outside the core, $N\sim \omega_0 r/R_*$ and $\mathfrak{g} \sim r \omega_0^2$, and the sound speed $c_s$ is nearly constant, with $c_s \sim \omega_0 R_*$. The radial displacement ${\bm \sigma}$ is regular near the center; in fact $\sigma_r/r$ is an analytic function everywhere including around the center $r \rightarrow 0$. From these observations, we can check that none of the potential terms given by Eq.~\eqref{eq:3ModeRad} pick up large contributions as we integrate through the star, and that their contribution is of the order $\sim {\cal O}(1)$ in the large frequency ratio $\omega_p/\omega_g$.

To sum up, we have shown that all the corrections to the frequency of the almost-neutrally stable $g$-mode due to inhomogeneous driving and the lowest nonlinear interactions are small. Specifically, they are the form $\epsilon$ or $\epsilon^2$ times terms of ${\cal O}(1)$ in the frequency ratio $\omega_p/\omega_g$. Since $\epsilon$ is a small parameter [$\epsilon = \Omega^2/\omega_0^2 = (m/M)(R_*/a)^3$, where $\Omega$ is the orbital frequency of the binary] during the early part of the in-spiral phase, the interaction of internal modes with the equilibrium tide does not cause them to go unstable in this region of parameter space.

\section{Discussion}
\label{sec:Discussion}

The volume-preserving coordinate transformation enables us to calculate the four-mode coupling terms which arise when we look at the interaction of the equilibrium tide with two high-order $p$- and $g$-modes, as long as we stay within the Cowling approximation. This coupling is important since we are interested in the effect of nonlinear interactions on the frequencies of the almost-neutrally stable $g$-modes, and for consistency we should consider all corrections which affect it at a given order in the tidal strength $\epsilon$. Using this estimation of the four-mode terms, we have found that there are no large corrections to $\omega_g$ up to the second order in $\epsilon$. This is true because of cancellations between large terms, some arising in the three-mode coupling and others in the four-mode coupling terms.

The cancellation occurs transparently using our method, but it is a useful check to see if the four-mode coupling computed using more traditional methods contains terms of the appropriate size for this cancellation. Using Eq.~(49) from~\citet{VanHoolst1994}, where repeated coordinate indices are summed over, we see that there are terms of the form 
\begin{align}
\kappa_{gg\chi^{(1)}\chi^{(1)}} &= - \frac{1}{3E_0} \int d^3 x  \, \Gamma_1 P \left[ (\nabla_{j} \chi^{(1)k}) (\nabla_{k} \xi_g^j ) \right]^2 + \dots 
 \sim \frac{1}{E_0} \int dr \, r^2 \Gamma_1 P \left( \frac{d g_h}{dr}\right)^2 
 \sim \frac{\omega_0^2 R_*^2  k_g^2 }{\omega_g^2} \sim \frac{\omega_p^2}{\omega_g^2}
\,.
\end{align}
For the first approximation, we have neglected factors of order unity, as well as lower order terms in the expression for $\kappa_{gg\chi \chi}$. For the second, we have used the simple stellar model and WKB eigenfunctions from \S\ref{sec:Compute}. The final relation uses the fact that $k_g \simeq k_p \simeq \omega_p/c_s \sim \omega_p/(\omega_0 R_*)$. We can see that this term has the right size to cancel with the square of the large three-mode term in equation for the perturbed frequency, Eq.~\eqref{eq:omegaminus}.

Although the analysis of this paper has introduced the four-mode corrections to the stability of the star, a number of assumptions and approximations have been invoked along the way. We now briefly summarize them, discuss their validity, and comment on topics for future investigation.

{\slshape Higher-order couplings}: Given that four-mode interactions have an important role in keeping the modes stable, one might wonder whether the five- and higher-mode couplings are important. The immediate answer to these concerns is that these couplings do not correct $\omega_g$ at ${\cal O}(\epsilon^2)$ -- the terms we have considered are the only terms which enter at this order. Of course, we have not shown that the star is stable against these sorts of non-resonant instabilities at even higher orders in $\epsilon$. However, our results and the intuition we derive from both our toy model and our coordinate transformation method indicate the reason that many large, cross-canceling terms enter into the analysis is that the mode expansion of the Lagrangian displacement $c_a$ is not the most natural choice of coordinates in the full nonlinear problem. We suspect that there exists some other coordinate system, in analogy to the simply rotated basis of the harmonic oscillator, which is better suited for analyzing the stability of the star, and in particular where the ``valleys'' of the potential energy surface (which would be exactly flat for a neutrally buoyant star with uniform specific entropy and composition) are straightened out. A more natural coordinate system could be related to our transformation, but we have not investigated this issue further. The search for a more natural coordinate system can serve as the subject of future work.

{\slshape Dynamical tide}: We have not considered the stability of the dynamical tide in this paper; it is not amenable to study using the volume-preserving transform here, which made use of the equilibrium nature of the background in an essential way. However, the dynamical tide is transiently excited during $\ell=2$ $g$-mode resonance crossings -- thus the energy input and gravitational wave template phase error do not depend on the details of the damping mechanism (see discussion in WAB \S5.3), and in any case the phase error is tiny \citep[Eq. 7.5]{Lai1994}.

{\slshape Cowling approximation}: The Cowling approximation, while a very good description of high-order $p$- and $g$-modes, is a poor model for the tidal bulge. The volume-preserving transformation makes use of the Cowling approximation in an essential way since we need to know the gravitational potential in order to construct it. This is however not a critical omission: since we are examining dynamics of only the daughter modes and not the excitation of the tidal bulge, we could have used as our background potential $U$ the perturbed potential including both the tidal bulge and the external field instead of just the latter. (This would have required including higher derivatives of the potential, e.g. $U_{abc}$, since then $U({\bm x})$ is not a quadratic function of ${\bm x}$, but these terms do not affect the arguments about the volume-preserving transform.)

{\slshape Time dependence of the external tidal field}: Throughout this study, we considered stability to a static perturbation of the star, since the instability in WAB exists even for a static perturbation (constant amplitude of the parent). If we instead consider the physically relevant scenario, where the tidal field is sourced by a distant companion in a circular orbit, the matrix of potential energies ${\cal M}$ for the daughter modes is positive-definite at any given time, but it varies at the orbital frequency, and one may wonder whether this leads to an instability. We investigate this in Appendix~\ref{app:Rotation}, and show that at second order in the tidal field, the {\em only} mathematically possible instabilities in this problem are the parametric resonance instability, the quasi-static instability considered by WAB and revisited here, and a centrifugal correction to the latter due to the rotation of ${\cal M}$ (as discussed in \S\ref{ss:toy}). 
The parametric resonance instability was considered in WAB and found not to occur for the equilibrium tide. The centrifugal instability would occur with a growth timescale of order
\begin{align}
t_{\rm cen} \sim \frac1{\epsilon\Omega} = \frac{\omega_0^2}{\Omega^3} \approx 4 \left( \frac{f_{\rm gw}}{100\rm\,Hz} \right)^{-3}\,{\rm s}
\end{align}
for parameters in WAB and $f_{\rm gw}\equiv 2\Omega$, and only for modes with $\omega_g\lesssim t_{\rm cen}^{-1}$. This is very slow compared to the original growth rate estimated in WAB, of order $\sim (\epsilon\omega_p)^{-1}$. Indeed, one may think of the centrifugal modification to the quasi-static instability as resulting from a failure at order $\Omega^2/\omega_p^2$ of the near-exact cancellation of three-mode and four-mode contributions to $\omega_-^2$. In principle a more detailed analysis would be required to determine whether the high-order $g$-modes can grow due to the centrifugal instability. However, the centrifugal instability timescale is comparable to the gravitational wave inspiral timescale (see WAB Eq. 20),
\begin{align}
\frac{t_{\rm cen}}{t_{\rm gw}} \sim 0.6 \left( \frac{M_{\rm chirp}}{1.2\,M_\odot} \right)^{5/3} \left( \frac{f_{\rm gw}}{100\,\rm Hz} \right)^{-1/3},
\end{align}
where $M_{\rm chirp} = \mu^{3/5} M_{\rm tot}^{2/5}$ is the chirp mass of the binary, $\mu$ here is the reduced mass of the binary, and $M_{\rm tot}$ is the total mass of the binary.
Thus the instability has time only to grow by of order one $e$-fold during the inspiral phase, and even this is neglecting any viscous damping of the $g$-modes. 
Detailed investigation of the factors of order unity and the possibility of modest growth due to centrifugal effects at the very latest stages of the inspiral is left to future work.

To summarize, we have shown that four-mode couplings play a critical role in the stability of equilibrium tides, and almost exactly cancel the three-mode coupling terms responsible for the $p$-mode-$g$-mode instability identified by WAB. This near-cancellation is generic and not dependent on the details of the equation of state. We conclude that in the quasi-static approximation the $p$-mode-$g$-mode instability and its deleterious effects on template-based searches for binary neutron stars go away. The principal caveat is that, when the time variability of the tidal field is taken into account, this near-cancellation is not exact, and it remains possible that an instability would develop over a much longer timespan -- but likely longer than the lifetime of the binary system.

The tools we have developed in this paper are quite general, and can be applied to a variety of interacting binary systems, including white dwarf binaries, stellar binaries, and possibly close planetary systems. A detailed treatment of the first nonlinear effects in stellar binaries may be needed to understand such systems, or very long-term secular effects in compact binaries.

White dwarf binaries are particularly well suited to studying tidal effects on inspirals. The physics describing tidal dissipation in these systems is much richer than in the case of neutron stars, primarily because these binaries inspiral for a much longer time in units of the dynamical timescale ($\omega_0^{-1}$). The dimensionless tidal strength $\epsilon$ evolves with time as
\begin{align}
  \frac{1}{\omega_0}\frac{d \ln{\epsilon}}{dt} \sim \epsilon^{4/3} (\omega_0 M_\textrm{chirp})^{5/3} \,.
\end{align}
The dynamical frequency goes as the square root of the density, and the density of a typical neutron star is around $10^{8}$ times that of a typical white dwarf. Hence, a white dwarf binary spends more time (in units of the dynamical timescale) at a given value of tidal strength than a neutron star binary, by a factor of $\sim 10^6$. For these slowly-evolving systems, subtler secular effects can become important, as well as both linear and nonlinear wave damping and the effect of entropy injection from tidal dissipation on the background state of the star \citep{Fuller11, Fuller12, Fuller13, Burkart12}. Similarly, the cumulative effect of tidal lag can result in the spin-up of the white dwarfs, which both ``Doppler shifts'' the tidal field to lower frequencies in the frame rotating with the white dwarf, and modifies the mode spectrum due to Coriolis forces \citep[e.g.][]{Bildsten96}. For the same reasons that they appeared here, four-mode couplings are likely to be necessary for a consistent treatment of nonlinear effects, and the results and techniques of this study may be of use for future lines of inquiry in these systems.

\acknowledgements

We thank Yanbei Chen for useful discussions regarding volume-preserving displacements. We thank Samaya Nissanke and Dave Tsang for useful discussions and comments on this manuscript. We also thank Anthony Piro for valuable comments on this manuscript. TV was supported by the International Fulbright Science and Technology Award. AZ was supported by NSF Grant PHY-1068881, CAREER Grant PHY-0956189, and by the David and Barbara Groce Startup Fund. CH was supported by the Simons Foundation, the David \& Lucile Packard Foundation, and the US Department of Energy (award DE-SC0006624).

\begin{appendix}

\section{Gravitational Potential Energy and the Cowling Approximation}
\label{sec:FieldDOF}

In this appendix, we briefly go into greater detail regarding the division of gravitational potential energy between the fields and the fluid elements, and the manner in which it relates to the Cowling approximation. The essential idea is that when writing the gravitational potential energy of a system of particles, there is an ambiguity in how to divide the energy between that stored in the gravitational field itself and that stored by the particles because of their position in the field. Here we follow the discussion presented in Chapter 13 of \cite{BlandfordThorne}. One can choose a constant $\beta$ and write the gravitational potential energy of a mass distribution $\rho$ as
\begin{align}
{\mathcal V}_{\rm grav}= \int d^3 x \left[(1 - \beta) \rho \Phi + (1 - 2 \beta) \frac{{\bm \nabla} \Phi \cdot {\bm \nabla} \Phi}{8 \pi G} \right]\,.
\end{align}
The Poisson equation then guarantees that any choice of $\beta$ gives the same total gravitational potential energy as any other. However, when we use the Cowling approximation, the Poisson equation is no longer valid; instead, the gravitational potential $\Phi$ is frozen to its value on the background matter distribution. As such, we need to take some care in our division of potential energy.

The usual choice for the constant $\beta$ when the self gravity of a system of particles is important is $\beta = 1/2$, since it places the gravitational potential energy into the pairwise interactions of the particles. The most natural choice when using the Cowling approximation is to choose $\beta = 0$, as this gives the usual expression for the gravitational potential energy of matter in an externally prescribed potential $\Phi$. The integral over the energy density of the field becomes our constant $\mathcal C$ in Eq.~\eqref{eq:GeneralV}, and this accounts for the lack of a factor of $1/2$ in front of the term $\Phi_0 + \epsilon U$ in these equations.

\section{Non-Axisymmetric Modes}
\label{sec:Nonaxisymm}

In this appendix, we extend our analysis to the full set of non-axisymmetric modes with all possible values of $m$.  As before, we consider a particular pair of $p$- and $g$-modes with angular momentum quantum numbers $l_p$ and $l_g$. Since these are coupled to the $l=2$ tidal potential $U$, the triad $(l_p, l_g, 2)$ satisfies the triangle inequality, and their sum is even. There are $2l_p+1$ $p$-modes with azimuthal quantum numbers $m_p$ ranging from $-l_p$ to $l_p$, and likewise a number of $g$-modes, whose unperturbed frequencies $\omega_p$ and $\omega_g$ are independent of $m$ by the symmetry of the background star. As before, we set $\omega_p > \omega_g$. The sub-block of $\mathcal M$ for this pair of modes is
\begin{align}
\mathcal{M} = &
  \begin{pmatrix}
     & \mathcal{M}_{pp} & &   & \mathcal{M}_{pg} & & \\
       &        &     &   &        \\
       &   \mathcal{M}_{gp}  & &  & \mathcal{M}_{gg} & & \\
\multicolumn{3}{c}{$\upbracefill$} & \multicolumn{3}{c}{$\upbracefill$}\\
\multicolumn{3}{c}{\scriptstyle 2l_p + 1} & \multicolumn{3}{c}{\scriptstyle 2l_g + 1}\\
\noalign{\vspace{-2\normalbaselineskip}}
  \end{pmatrix},
\end{align}
where
\begin{align}
\noalign{\vspace{2\normalbaselineskip}}  
[\mathcal M_{pp}]_{m_1,m_2} = & \omega_p^2 \left[ \delta_{m_1,m_2} - \epsilon \left( U_{\bar{p}_{m_1} p_{m_2}} + \sum 2 \kappa_{a \bar{p}_{m_1} p_{m_2}} \chi_a^{(1)} \right) - \epsilon^2 \sum \left( 2 \kappa_{a \bar{p}_{m_1} p_{m_2}} \chi_a^{(2)} + 3 \kappa_{a b \bar{p}_{m_1} p_{m_2}} \chi_a^{(1)} \chi_b^{(1)} \right)\right] \nonumber \\&+ {\cal O}(\epsilon^3), \nonumber\\
[\mathcal M_{pg}]_{m_1,m_2} = & [\mathcal M_{gp}]_{m_2,m_1}^{*} = \omega_p \omega_g \left[ -\epsilon \left(U_{\bar{p}_{m_1} g_{m_2}} + \sum 2 \kappa_{a \bar{p}_{m_1} g_{m_2}} \chi_a^{(1)} \right) \right]+ {\cal O}(\epsilon^2)  \,,{\rm~~and}\nonumber \\
[\mathcal M_{gg}]_{m_1,m_2} = &  \omega_g^2 \left[ \delta_{m_1,m_2} - \epsilon \left( U_{\bar{g}_{m_1} g_{m_2}} + \sum 2 \kappa_{a \bar{g}_{m_1} g_{m_2}} \chi_a^{(1)} \right) - \epsilon^2 \sum \left( 2 \kappa_{a \bar{g}_{m_1} g_{m_2}} \chi_a^{(2)} + 3 \kappa_{a b \bar{g}_{m_1} g_{m_2}} \chi_a^{(1)} \chi_b^{(1)} \right)\right]\nonumber \\& + {\cal O}(\epsilon^3)  \,.
\end{align} 
In the above expressions, the convention is that $\mathcal M_{a b}$ is the coefficient of $\eta_{a}^{\prime}{^*}\, \eta'_{b}$ in the expansion of the Lagrangian given by Eq.~\eqref{eq:etaL}, after the rescaling $\eta'_a = \eta_a/\omega_a$. In writing this expression, we have used the symmetry of the coupling coefficients and the reality of ${\bm \chi}$.

\begin{figure*}
\centering
\includegraphics[width=2.25in]{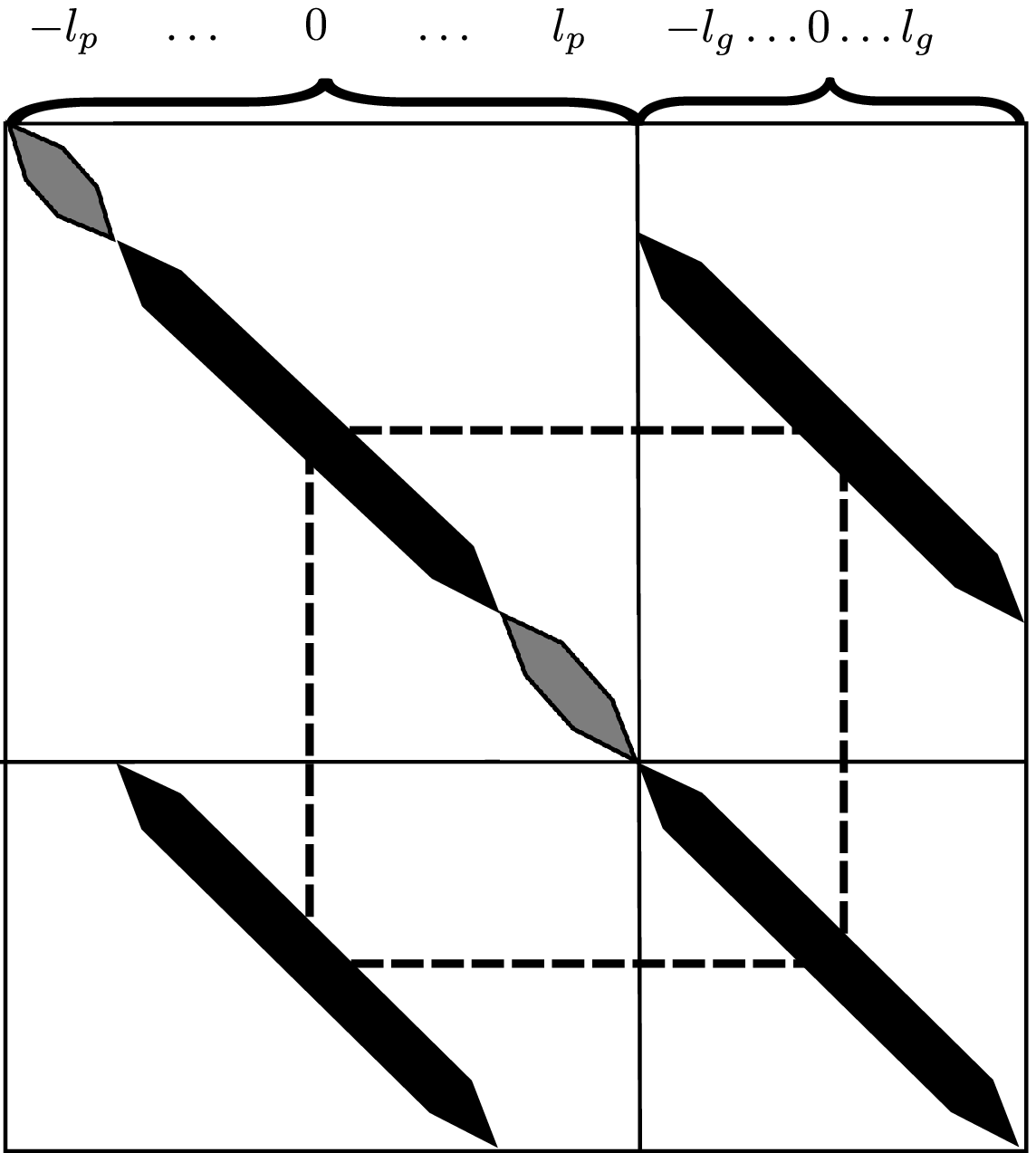} \hspace{0.5in}
\includegraphics[width=2.25in]{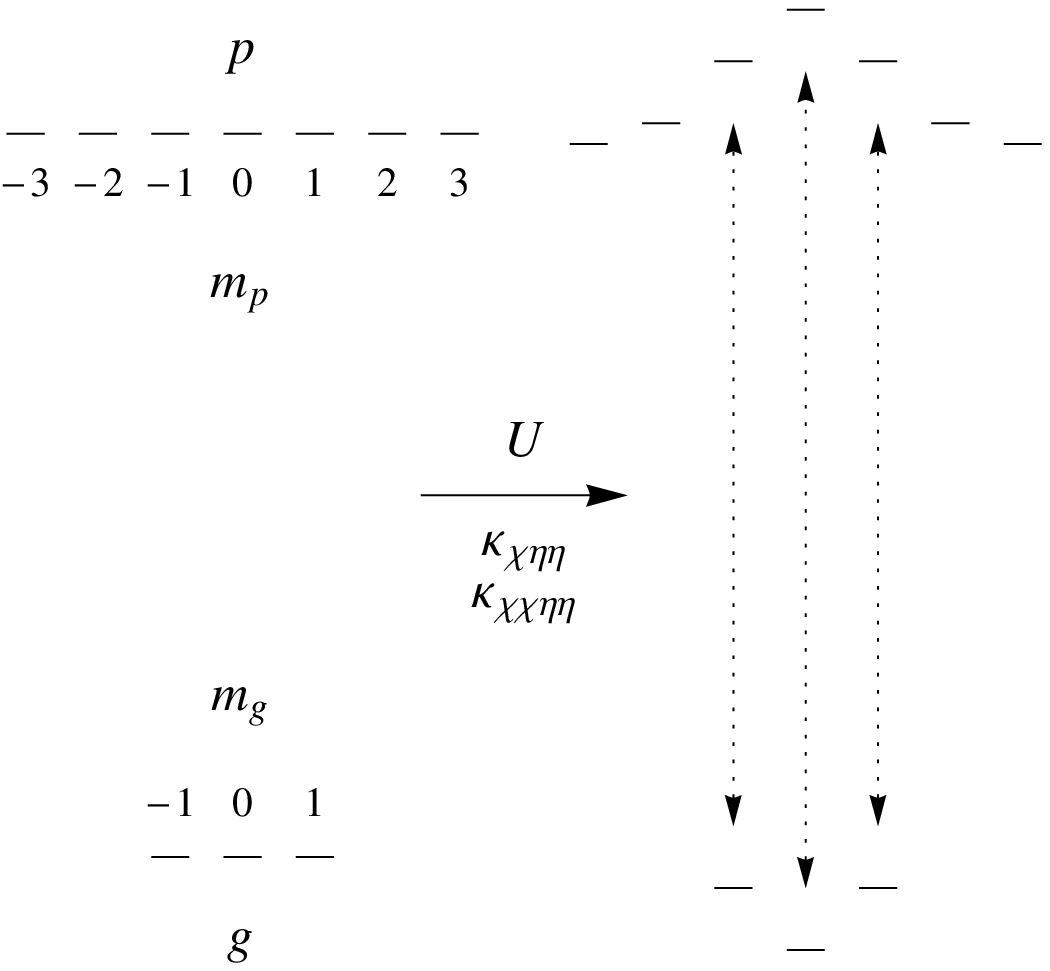}
\caption{Left panel: Illustration of the form of the sub-block of $\mathcal M$, for the interaction of two modes $p$ and $g$. We take $l_p> l_g$. Only the diagonals and those entries which couple terms with $m_p = m_g$ are nonzero. The entries marked with the black, thick line can be rearranged into independent $2\times2$ blocks, and an example block is marked out by the dashed square. Those entries of the sub-block which do not interact are marked with the gray thick line, and they form independent $1\times1$ blocks. Right panel: The tidal field and mode couplings correct the frequencies of the various modes as illustrated for a pair $(p,g)$ with $l_p=3$ and $l_g=1$. These corrections are due to both off-diagonal and diagonal perturbations in the sub-block of $\mathcal M$.}
\label{fig:Matrix}
\end{figure*}

Until now, we have not used the fact that $U$ and ${\bm \chi}$ are axisymmetric, i.e. they only have $m=0$ components. By angular momentum conservation, such a $U$ and ${\bm \chi}$ can only couple modes of the same azimuthal quantum number $m$. In addition to this, the strengths of the couplings between modes with $m$ are the same as those between modes with $-m$ by parity. This implies that the matrix $\mathcal{M}$ has a sparse structure as shown in the left hand side of Fig.~\ref{fig:Matrix}. In order to study the perturbations to a particular mode at the lowest order, we need to consider a $2 \times 2$ sub-matrix corresponding to the members of the $p$ and $g$ subspaces with the same quantum number $m$, when possible. Otherwise, the subspace is a $1\times1$ block.

The resulting perturbations to the mode frequencies are, to second order in $\epsilon$,
\begin{align}
  \omega_{p_m}^2 = &\, \omega_p^2 \left[ 1 - \epsilon \left( U_{\bar{p}_m p_m} + \sum 2 \kappa_{ a \bar{p}_m p_m} \chi_a^{(1)} \right) 
  - \epsilon^2 \sum\left(2 \kappa_{a \bar{p}_m p_m} \chi_a^{(2)} + 3 \kappa_{a b \bar{p}_m p_m} \chi_a^{(1)} \chi_b^{(1)} \right) \right] \nonumber \\
   & + \epsilon^2 \frac{ \omega_p^2 \omega_g^2 }{\omega_p^2 - \omega_g^2} \left| U_{\bar{p}_m g_m} + \sum 2 \kappa_{ a \bar{p}_m g_m} \chi_a^{(1)} \right|^2 {\rm~~and}\nonumber\\
  \omega_{g_m}^2 = &\, \omega_g^2 \left[ 1 - \epsilon \left( U_{\bar{g}_m g_m} + \sum 2 \kappa_{a \bar{g}_m g_m} \chi_a^{(1)} \right) 
  - \epsilon^2 \sum\left(2 \kappa_{a \bar{g}_m g_m} \chi_a^{(2)} + 3 \kappa_{a b \bar{g}_m g_m} \chi_a^{(1)} \chi_b^{(1)} \right) \right] \nonumber \\
   & - \epsilon^2 \frac{ \omega_p^2 \omega_g^2 }{\omega_p^2 - \omega_g^2} \left| U_{\bar{p}_m g_m} + \sum 2 \kappa_{a \bar{p}_m g_m} \chi_a^{(1)} \right|^2 .
\end{align}
We can recognize Eqs.~\eqref{eq:omegaplus} and \eqref{eq:omegaminus} as special cases of this for $m=0$. This level splitting is schematically shown in the right hand side of Fig.~\ref{fig:Matrix}. For any $m$, the analysis of the cancellations between the four-mode terms and the three-mode terms proceeds exactly as for the axisymmetric case.

\section{Reversing the Flow of an Infinitesimal Coordinate Transform}
\label{sec:Reverse}

In this appendix, we briefly derive the identity for reversing the direction of the flow discussed in \S\ref{sec:VTransform}. Consider a point $\mathcal P$ on a manifold, and define an origin so that this point has the coordinate vector ${\bm x}_{\mathcal P}$. Next, consider the diffeomorphism generated by a flow in the direction of an infinitesimal vector field ${\bm \zeta}({\bm x})$. Viewed from the perspective of an active transformation, this diffeomorphism sends the points on the manifold a small distance along the integral curves of ${\bm \zeta}$, and the point $\mathcal P$ is mapped to a point $\mathcal Q$ whose coordinate vector is given by (keeping track of terms to second order in the small motion)
\begin{align}
\label{eq:ForwardMotion}
{\bm x}_{\mathcal Q} = {\bm x}_{\mathcal P} + {\bm \zeta}({\bm x}_{\mathcal P}) +\left. \frac 12 ({\bm \zeta} \cdot {\bm \nabla}) {\bm \zeta}\right|_{{\bm x}_{\mathcal P}} + \hdots \,.
\end{align}
We can then express ${\bm x}_{\mathcal P}$ as
\begin{align}
{\bm x}_{\mathcal P} & = {\bm x}_{\mathcal Q} - {\bm \zeta}({\bm x}_{\mathcal P}) -\left. \frac 12 ({\bm \zeta} \cdot {\bm \nabla}) {\bm \zeta}\right|_{{\bm x}_{\mathcal P}} + \hdots  =  {\bm x}_{\mathcal Q} - {\bm \zeta}({\bm x}_{\mathcal Q}) + \left. ({\bm \zeta} \cdot {\bm \nabla}) {\bm \zeta}\right|_{{\bm x}_{\mathcal Q}} -\left. \frac 12 ({\bm \zeta} \cdot {\bm \nabla}) {\bm \zeta}\right|_{{\bm x}_{\mathcal Q}} + \hdots \notag \\
& =  {\bm x}_{\mathcal Q} - {\bm \zeta}({\bm x}_{\mathcal Q}) + \frac 12 \left. ({\bm \zeta} \cdot {\bm \nabla}) {\bm \zeta}\right|_{{\bm x}_{\mathcal Q}} + \hdots \,,
\end{align}
where we have used Eq.~\eqref{eq:ForwardMotion} to eliminate ${\bm x}_{\mathcal P}$ from the right side of the equation. On the other hand, we could have considered the inverse flow, carrying ${\bm x}_{\mathcal Q}$ to ${\bm x}_{\mathcal P}$. It is clear that this flow is accomplished by simply reversing the sign of the generator field ${\bm \zeta}$ and evaluating it at the new base point ${\bm x}_{\mathcal Q}$. Generalizing to the coordinates of points on the entire manifold, if we call the coordinates we begin with ${\bm x}$ and transform to coordinates ${\bm X}$ by the infinitesimal transform ${\bm \zeta}$, the transformation is
\begin{align}
{\bm X} =  {\bm x} + {\bm \zeta}({\bm x}) + \left. \frac 12 ({\bm \zeta} \cdot {\bm \nabla}) {\bm \zeta}\right|_{\bm x} + \hdots \,,
\end{align}
while the inverse transform is 
\begin{align}
{\bm x}  =  {\bm X}- {\bm \zeta}({\bm X}) + \left. \frac 12 ({\bm \zeta} \cdot {\bm \nabla}) {\bm \zeta}\right|_{\bm X} + \hdots \,.
\end{align}
These relations allow us to invert our coordinate transform derived in \S\ref{sec:VTransform}.

\section{Additional Details in Estimating the Perturbations to the Eigenfrequencies}
\label{sec:CalcDetails}

Here, we collect the details of the various computations needed to estimate the various terms in Eq.~\eqref{eq:omegaM2}. As before we define $\Lambda_a^2=l_a(l_a+1)$. The angular integrals are
\begin{align}
T_{abc} = & \int d \Omega \, Y_a Y_b  Y_c = \sqrt{\frac{(2l_a+1)(2l_b+1)(2l_c+1)}{4 \pi} } 
\left(
\begin{array}{ccc}
l_a & l_b & l_c \\
m_a & m_b & m_c
\end{array}
\right)
\left(
\begin{array}{ccc}
l_a & l_b & l_c \\
0 & 0 & 0
\end{array}
\right)\,, \\
F_{a,bc}  = & \int d \Omega \, Y_a ({\bm \nabla} Y_b)\cdot({\bm \nabla} Y_c) = \frac {T_{abc}}{2} (\Lambda_b^2 + \Lambda_c^2 - \Lambda_a^2 ) \,, \\
G_{a,bc}  = & \int d \Omega \, (\nabla_j \nabla_k Y_a) (\nabla^j Y_b) (\nabla^k Y_c) = \frac{T_{abc}}{4} \left[\Lambda_a^4 -( \Lambda_b^2 - \Lambda_c^2)^2 \right] \,, \\
S_{abc} = & \frac{1}{2} (\Lambda_a^2 F_{a,bc} + \Lambda_b^2 F_{b,ca} + \Lambda_c^2 F_{c,ab}) \,, {\rm~~and}\\
V_{a, bc} = & \Lambda_b^2 \Lambda_c^2 T_{abc} - F_{a,bc} - S_{abc} \,,
\end{align}
where for $G_{a,bc}$ we have used Einstein summation convention and we use the standard metric on a unit sphere to raise indices. 

In the remainder of this section, we expand on our derivation of the three-mode coupling $\kappa_{\sigma gg}$, with two of the modes being high-order $g$-modes and the third being the radial displacement ${\bm \sigma}$. As mentioned in subsection \ref{sec:RemTerms}, we build upon the results of \citet{Weinberg2012}. We take as our starting point their Eqs.~(A27)--(A30), 
\begin{align}
  \kappa_{abc} = & \frac{1}{2E_0} \int dr \left[ 
    r^2 P \left\{ \Gamma_1(\Gamma_1 - 2) + \frac{\partial \Gamma_1}{\partial \ln \rho} \Big\rfloor_s \right\} ({\bm \nabla} \cdot {\bm a})_r ({\bm \nabla} \cdot {\bm b})_r ({\bm \nabla}\cdot {\bm c})_r T_{abc} 
      \right.  \label{eq:e1l1}\\ 
      & \hspace{10pt} \left. 
      + 2T_{abc} \rho \mathfrak{g} a_r b_r c_r 
      + (F_{a,bc}+S_{abc}) \rho \mathfrak{g} a_r b_h c_h
      + (F_{b,ca}+S_{abc}) \rho \mathfrak{g} b_r c_h a_h
      + (F_{c,ab}+S_{abc}) \rho \mathfrak{g} c_r a_h b_h
      \right. \nonumber \\
      & \hspace{10pt} \left.
      - T_{abc} \Lambda_a^2 \rho \mathfrak{g} a_h b_r c_r
      - T_{abc} \Lambda_b^2 \rho \mathfrak{g} b_h c_r a_r
      - T_{abc} \Lambda_c^2 \rho \mathfrak{g} c_h a_r b_r
      - 2S_{abc}\rho \mathfrak{g} a_h b_h c_h
       \label{eq:e1l2} \right. \\
      & \hspace{10pt} \left.
      + \left( \frac{da_r}{dr} \frac{db_r}{dr} + \frac{2}{r^2} a_r b_r \right) r^2 \Gamma_1 P ({\bm \nabla} \cdot {\bm c})_r T_{abc}
      + \left( a_r \frac{db_h}{dr} + b_r \frac{da_h}{dr} - \frac{d}{dr} \left( a_h b_h \right) \right)r \Gamma_1 P ({\bm \nabla} \cdot {\bm c})_r F_{c,ab}
      \right. \nonumber \\
      & \hspace{10pt} \left.
      - \left( \Lambda_b^2 a_r b_h + \Lambda_a^2 a_h b_r \right) \Gamma_1 P ({\bm \nabla} \cdot {\bm c})_r T_{abc}
      + a_h b_h \Gamma_1 P ({\bm \nabla} \cdot {\bm c})_r V_{c,ab}
      \right. \nonumber \\
      & \hspace{10pt} \left.
      + \left( \frac{db_r}{dr} \frac{dc_r}{dr} + \frac{2}{r^2} b_r c_r \right) r^2 \Gamma_1 P ({\bm \nabla} \cdot {\bm a})_r T_{abc}
      + \left( b_r \frac{dc_h}{dr} + c_r \frac{db_h}{dr} - \frac{d}{dr} \left( b_h c_h \right) \right)r \Gamma_1 P ({\bm \nabla} \cdot {\bm a})_r F_{a,bc}
      \right. \nonumber \\
      & \hspace{10pt} \left.
      - \left( \Lambda_c^2 b_r c_h + \Lambda_b^2 b_h c_r \right) \Gamma_1 P ({\bm \nabla} \cdot {\bm a})_r T_{abc}
      + b_h c_h \Gamma_1 P ({\bm \nabla} \cdot {\bm a})_r V_{a,bc}
      \right. \nonumber \\
      & \hspace{10pt} \left.
      + \left( \frac{dc_r}{dr} \frac{da_r}{dr} + \frac{2}{r^2} c_r a_r \right) r^2 \Gamma_1 P ({\bm \nabla} \cdot {\bm b})_r T_{abc}
      + \left( c_r \frac{da_h}{dr} + a_r \frac{dc_h}{dr} - \frac{d}{dr} \left( c_h a_h \right) \right) r \Gamma_1 P ({\bm \nabla} \cdot {\bm b})_r F_{b,ca}
      \right. \nonumber \\
      & \hspace{10pt} \left.
      - \left( \Lambda_a^2 c_r a_h + \Lambda_c^2 c_h a_r \right) \Gamma_1 P ({\bm \nabla} \cdot {\bm b})_r T_{abc}
      + c_h a_h \Gamma_1 P ({\bm \nabla} \cdot {\bm b})_r V_{b,ca}
       \label{eq:e1l3} \right. \\
      & \hspace{10pt} \left.
      - r^2 \rho a_r b_r c_r \frac{d^2\mathfrak{g}}{dr^2} T_{abc}
      - r^2 \rho \frac{d}{dr}\left(\frac{\mathfrak{g}}{r}\right) (F_{a,bc} a_r b_h c_h + F_{b,ca} a_h b_r c_h + F_{c,ab} a_h b_h c_r) \right] \,.
       \label{eq:e1l4}
\end{align}
In writing this, we are including only those terms which arise within the Cowling approximation. The above expression assumes that the radial mode, taken to be the ${\bm a}$ mode, is defined in the manner of Eq.~\eqref{eq:chiexpansion}, i.e. its radial component $a_r$ is such that ${\bm a} = a_r Y_{00}(\theta,\phi) \hat{r}$. As such, when we later plug in the radial displacement ${\bm \sigma}$, we need to account for our different normalization by inserting stray factors of $\sqrt{4\pi}$. We consider the case where the other two modes are azimuthally symmetric $g$-modes ($m=0$). Relaxing this assumption leads to extra phase factors of $(-1)^{-m}$, which do not affect our conclusions.

In order to simplify these terms, we initially proceed as in \citet{Weinberg2012}. We integrate by parts the $d(a_h b_h)/dr$ terms in line \eqref{eq:e1l3} and cancel against the $S_{abc} \rho \mathfrak{g} a_h b_h c_h $ term in line \eqref{eq:e1l2} after using the radial equation of motion~\eqref{eq:ModeRad} for the displacements. We then deal with the $a_r \, db_h/dr$ terms in the same manner. Next, we use the definition of divergence~\eqref{eq:ModeDiv} for the modes to eliminate factors of $da_r/dr$ in line \eqref{eq:e1l2}, and residual terms which are leftover from the previous steps. However, we cannot subsequently use the angular equations of motion to simplify terms with one factor of $({\bm \nabla} \cdot {\bm a})_r$, as one of the modes is radial.

When we use the radial equation of motion~\eqref{eq:ModeRad} for the modes, we collect the restoring force terms into a homogeneous part $\kappa_{abc,H}$ and the driving terms into an inhomogeneous part $\kappa_{abc,I}$. The homogeneous part is given by
\begin{align}
  \kappa_{abc,H} = & \frac{1}{2E_0} \int dr \left[ 
    r^2 P \left\{ \Gamma_1(\Gamma_1 + 1) + \frac{\partial \Gamma_1}{\partial \ln \rho} \Big\rfloor_s \right\} ({\bm \nabla} \cdot {\bm a})_r ({\bm \nabla} \cdot {\bm b})_r ({\bm \nabla}\cdot {\bm c})_r T_{abc} 
    \right. \label{eq:e2l1}\\
    & \hspace{10pt} \left.
    + \Gamma_1 P r ({\bm \nabla} \cdot {\bm a})_r ({\bm \nabla} \cdot {\bm c})_r \left\{ - 4 b_r + \Lambda_b^2 b_h \right\} T_{abc}
    + \Gamma_1 P r ({\bm \nabla} \cdot {\bm b})_r ({\bm \nabla} \cdot {\bm a})_r \left\{ - 4 c_r + \Lambda_c^2 c_h \right\} T_{abc}
    \right. \nonumber \\
    & \hspace{10pt} \left.
    + \Gamma_1 P r ({\bm \nabla} \cdot {\bm c})_r ({\bm \nabla} \cdot {\bm b})_r \left\{ - 4 a_r + \Lambda_a^2 a_h \right\} T_{abc}
    \right. \label{eq:e2l2} \\
    & \hspace{10pt} \left.
    + \Gamma_1 P ({\bm \nabla} \cdot {\bm c})_r \left( G_{c, ab} a_h b_h + 6 T_{abc} a_r b_r + \left\{ F_{c, ab} - 3 \Lambda_b^2 T_{abc} \right\} a_r b_h + \left\{ F_{c, ab} - 3 \Lambda_a^2 T_{abc} \right\} b_r a_h \right)
    \right. \nonumber \\
    & \hspace{10pt} \left.
    + \Gamma_1 P ({\bm \nabla} \cdot {\bm a})_r \left( G_{a, bc} b_h c_h + 6 T_{abc} b_r c_r + \left\{ F_{a, bc} - 3 \Lambda_c^2 T_{abc} \right\} b_r c_h + \left\{ F_{a, bc} - 3 \Lambda_b^2 T_{abc} \right\} c_r b_h \right)
    \right. \nonumber \\
    & \hspace{10pt} \left.
    + \Gamma_1 P ({\bm \nabla} \cdot {\bm b})_r \left( G_{b, ca} a_h c_h + 6 T_{abc} c_r a_r + \left\{ F_{b, ca} - 3 \Lambda_a^2 T_{abc} \right\} c_r a_h + \left\{ F_{b, ca} - 3 \Lambda_c^2 T_{abc} \right\} a_r c_h \right)
    \right. \label{eq:e2l3} \\
    & \hspace{10pt} \left. 
    + T_{abc} \left( 2 \mathfrak{g} - r^2 \frac{d^2 \mathfrak{g}}{d r^2} \right) \rho  a_r b_r c_r 
    + S_{abc} \rho \mathfrak{g} ( a_r b_h c_h + b_r c_h a_h + c_r a_h b_h )
    \right. \nonumber \\
    & \hspace{10pt} \left.
    + \rho \left( \mathfrak{g} - r \frac{d \mathfrak{g}}{d r} \right) \left\{ \Lambda_a^2 a_h b_r c_r + \Lambda_b^2 b_h c_r a_r + \Lambda_c^2 c_h a_r b_r \right\} T_{abc}
    \right. \nonumber \\
    & \hspace{10pt} \left.
    - \rho r \left( F_{c, ab} \omega_c^2 a_h b_h c_r + F_{a, bc} \omega_a^2 a_r b_h c_h + F_{b, ca} \omega_b^2 a_h b_r c_h \right) 
    - (a_r b_h + a_h b_r ) F_{c, ab} \left( \Lambda_c^2 \rho \mathfrak{g} c_h - \omega_c^2 \rho r c_r \right)
    \right. \nonumber \\
    & \hspace{10pt} \left.
    - (b_r c_h + b_h c_r ) F_{a, bc} \left( \Lambda_a^2 \rho \mathfrak{g} a_h - \omega_a^2 \rho r a_r \right)
    - (c_r a_h + c_h a_r ) F_{b, ca} \left( \Lambda_b^2 \rho \mathfrak{g} b_h - \omega_b^2 \rho r b_r \right)
    \right] \,.
\end{align}
For a triplet of modes $(\sigma, g, g)$ with angular quantum numbers $(\{0,0\},\{l_g,0\},\{l_g,0\})$, the angular integrals in the above expression are
\begin{align}
  T_{\sigma gg} = & \sqrt{\frac{1}{4\pi}}\,, &  (F_{\sigma, gg},F_{g,g\sigma}) = & T_{\sigma gg}(\Lambda_g^2,0) \,, &
  (G_{\sigma, gg},G_{g, \sigma g}) = & (0,0)\,, &{\rm and~~~~~~}
  S_{\sigma gg} = & 0\,.
\end{align}
Plugging in the forms of the displacements, the angular integrals, and the fact that $\omega_\sigma = 0$, we arrive at the expression
\begin{align}
  \epsilon^2 \kappa_{\sigma gg,H} = & \frac{1}{2E_0} \int dr \left[ 
    r^2 P \left\{ \Gamma_1(\Gamma_1 + 1) + \frac{\partial \Gamma_1}{\partial \ln \rho} \Big\rfloor_s \right\} ({\bm \nabla} \cdot {\bm \sigma}) ({\bm \nabla} \cdot {\bm g})_r ({\bm \nabla}\cdot {\bm g})_r
      \right. \nonumber\\
      & \hspace{10pt} \left.
      + 2 \Gamma_1 P r ({\bm \nabla} \cdot {\bm \sigma}) ({\bm \nabla} \cdot {\bm g})_r \left\{ - 4 g_r + \Lambda_g^2 g_h \right\}
      - 4 \Gamma_1 P r ({\bm \nabla} \cdot {\bm g})_r ({\bm \nabla} \cdot {\bm g})_r \sigma_r
      \right. \nonumber\\
      & \hspace{10pt} \left.
      + 6 \Gamma_1 P ({\bm \nabla} \cdot {\bm g})_r \sigma_r ( 2 g_r - \Lambda_g^2 g_h ) 
      + 2 \Gamma_1 P ({\bm \nabla} \cdot {\bm \sigma}) g_r ( 3 g_r - 2 \Lambda_g^2 g_h )
      \right. \nonumber\\
      & \hspace{10pt} \left. 
      + \rho \left( 2 \mathfrak{g} - r^2 \frac{d^2 \mathfrak{g}}{d r^2} \right) \sigma_r g_r g_r
      + \rho \left( 2 \mathfrak{g} - 2 r \frac{d \mathfrak{g}}{d r} \right) \Lambda_g^2 \sigma_r g_r g_h
      \right] \,.
\end{align}

As we have noted in \S\ref{sec:RemTerms}, the potential $V({\bm r})$ only couples to the radial displacement ${\bm \sigma}$ due to its spherical symmetry. Hence only the radial displacement has an inhomogeneous term in its equation of motion. Keeping track of when we have used the radial equation of motion (force-balance) for this displacement, this residual inhomogeneous term is
\begin{align}
  \label{eq:kappaggsigmaI}
\epsilon^2 \kappa_{\sigma gg,I} = & \frac{1}{2E_0} \int dr \Lambda_g^2 \rho r \left( g_h g_h - 2 g_r g_h \right) \epsilon^2 \frac{d V}{d r} \,.
\end{align}
These homogeneous and inhomogeneous contributions to the three-mode coupling, when combined with the nonlinear driving term from the potential as $\epsilon^2 (V_{gg} + 2\kappa_{\sigma gg, H} + 2\kappa_{\sigma gg,I})$, give the perturbation to the restoring force. Note that the largest contribution to this sum, due to the horizontal displacement of the $g$-mode in the two inhomogeneous terms given by Eq.~\eqref{eq:Vgg} and Eq.~\eqref{eq:kappaggsigmaI}, cancels exactly.

The remaining terms can be simplified further using the divergence equation, the radial equations of motion for the modes and the angular equation of motion for the $g$-mode. We have chosen to reduce them to the form given in Eq.~\eqref{eq:3ModeRad} in order to emphasize terms whose dominant dependence on the frequency ratio $\omega_p/\omega_g$ can be easily estimated from the WKB form of the $g$-mode eigenfunction. 
In particular, the only remaining term involving the horizontal displacement in Eq.~\eqref{eq:3ModeRad} contains the combination of factors $\omega_g^2 g_h g_h$, so that the small factor $\omega_g^2$ suppresses the large contribution from $g_h g_h$.

\section{Rotating tidal fields}
\label{app:Rotation}

We have thus far considered the stability of small daughter perturbations ${\bm\eta}$ for a static tidal field. This section considers a rotating tidal field, as occurs for a binary on a circular orbit, and shows that -- as we would expect -- the quasi-static instability (considered by WAB) and the parametric resonance instability are the only two possible instability mechanisms at second order in the tidal perturbation and while considering the behavior of two roots of the secular equation at a time. (The ``collective instability'' in \citealt{Weinberg2012} is of higher order in the sense that it requires multiple resonance criteria to be satisfied.) The only difference is that the quasi-static instability criterion is modified by the time dependence of the tidal field: the natural frequency of the $g$-mode oscillation $\omega_g^2$ picks up not just the static three- and four-mode coupling corrections, but a ``centrifugal'' correction due to the time variation of the shallow direction in configuration space (i.e. the small eigenvalue of ${\cal M}$).

We suppose that the daughter perturbation modes have an evolution matrix ${\cal M}$ in the instantaneous frame with the tidal field aligned on the $z$-axis, so that in the case of a static tidal field we have as before $\ddot{\bm\eta} = -{\cal M}{\bm\eta}$. (In the case where the tidal field is rotating, the frame where the tidal field is aligned with the $z$-axis is not an inertial frame, and the equations of motion have additional centrifugal and Coriolis corrections.)
The eigenvalues of ${\cal M}$ are the squares of the mode frequencies in the static approximation. Here ${\cal M}$ is an $N\times N$ matrix, where $N=(2\ell_p+1)+(2\ell_g+1)$ is the number of normal modes under consideration. In terms of the basis of Appendix~\ref{sec:Nonaxisymm}, ${\cal M}$ is symmetric, real, has nonzero entries only when the azimuthal quantum numbers are equal, and has equivalent components for $m$ and $-m$. (In the main text, we treated ${\cal M}$ as a $2\times 2$ matrix, since it is trivially block-diagonal by rotational symmetry and hence there is no need to consider more than 2 modes at the same time. For a rotating tidal field, we must generalize this.)

We would like to express ${\cal M}$ in an inertial frame where the binary orbit is in the $xy$-plane; this is achieved via the rotation by $\Omega t$ around the $z$-axis, followed by rotation by $\pi/2$ around the $y$-axis:
\begin{align}
{\bf R}(t) = \exp\left( - i\frac\pi2{\bf L}_y\right) \exp(-i\Omega t {\bf L}_z) = {\bf R}(0)
\exp(-i\Omega t {\bf L}_z),
\label{eq:E-rdef}
\end{align}
where ${\bf L}_x$, ${\bf L}_y$, and ${\bf L}_z$ are the angular momentum operators, each of which is an $N\times N$ Hermitian matrix. This is a unitary rotation matrix in the sense that if ${\bm\eta}^{\rm(inert)}$ is the daughter perturbation in the inertial frame, then ${\bm\eta} = {\bf R}(t){\bm\eta}^{\rm(inert)}$ is the daughter in the frame aligned with the instantaneous tidal field, where $\mathcal M$ is defined. The evolution equation in the rotated frame is
\begin{align}
\ddot{\bm\eta}^{\rm(inert)} = -{\bf R}^\dagger{\cal M}{\bf R}{\bm\eta}^{\rm(inert)}.
\label{eq:rotated}
\end{align}
We now define a partially rotated daughter perturbation ${\bm\mu} \equiv \exp(-i\Omega t {\bf L}_z){\bm\eta}^{\rm(inert)}$, in a basis where the $z$ axis is normal to the orbital plane, but the $x$ axis always points in the direction of the companion. 
We then left-multiply Eq.~(\ref{eq:rotated}) by $\exp(-i\Omega t {\bf L}_z)$ to get
\begin{align}
\exp(-i\Omega t {\bf L}_z)\frac{d^2}{dt^2}[\exp(i\Omega t {\bf L}_z) {\bm\mu}] = -{\bf R}^\dagger(0){\cal M}{\bf R}(0) {\bm\mu},
\end{align}
which expands to
\begin{align}
\ddot{\bm\mu} + 2i\Omega{\bf L}_z \dot{\bm\mu} - \Omega^2{\bf L}_z^2 {\bm\mu} = -{\bf R}^\dagger(0) {\cal M}{\bf R}(0) {\bm\mu}.
\label{eq:etemp}
\end{align}
It is convenient to Taylor-expand ${\cal M}$ in the tidal deformation,
\begin{align}
{\cal M} = {\cal M}^{(0)} + \delta{\cal M},{\rm~~~where~~~}
\delta{\cal M} = \epsilon{\cal M}^{(1)} + \epsilon^2{\cal M}^{(2)} + ...;
\end{align}
then since ${\cal M}^{(0)}$ is spherically symmetric, we have ${\bf R}^\dagger(0) {\cal M}^{(0)}{\bf R}(0)={\cal M}^{(0)}$.
If we substitute into Eq.~(\ref{eq:etemp}) and take a solution of the form ${\bm\mu}\propto e^{-i\omega t}$, then 
\begin{align}
{\bf A}(\omega){\bm\mu} \equiv [-\omega^2 {\bf I} + 2\omega\Omega{\bf L}_z - \Omega^2{\bf L}_z^2 + {\cal M}^{(0)} + {\bf R}^\dagger(0) \delta{\cal M}{\bf R}(0) ]{\bm\mu} = 0,
\label{eq:ev}
\end{align}
where ${\bf I}$ is the identity matrix. An unstable daughter can then exist if Eq.~(\ref{eq:ev}) has a nontrivial solution (i.e. ${\bm\mu}\neq0$) in the upper-half complex plane. Since $\det{\bf A}(\omega)$ is a $2N^{\rm th}$ order polynomial in $\omega$, there are $2N$ solutions. Also ${\bf L}_z$ and $\delta{\cal M}$ are real and symmetric, and ${\bf R}(0)$ is real (this is because in the standard basis, the generator ${\bf L}_y$ is purely imaginary; hence the finite rotation matrices around the $y$-axis have all real entries). Therefore $\det{\bf A}(\omega)$ is symmetric and has real coefficients, and so non-real solutions for $\omega$ occur in conjugate pairs.

It is also important to note that if we define the reflection matrix ${\bf \Sigma}$ through the $xz$-plane, so that in the spherical harmonic basis $\Sigma_{m,m'}=(-1)^m\delta_{m,-m'}$, we have that ${\bf \Sigma}$ anti-commutes with ${\bf L}_z$, i.e. $\{{\bf \Sigma},{\bf L}_z\}=0$, but ${\bf \Sigma}$ commutes with ${\cal M}$ and ${\bf R}(0)$. It follows that ${\bf \Sigma}{\bf A}(\omega) = {\bf A}(-\omega){\bf \Sigma}$ and, since ${\bf \Sigma}$ is nonsingular, $\det {\bf A}(\omega)=\det{\bf A}(-\omega)$. Thus if $\omega$ is a solution, then so is $-\omega$.

In the {\em unperturbed} case where $\delta{\cal M}=0$, we can immediately see that ${\bf A}(\omega)={\bf A}^{(0)}(\omega)$ is diagonal in the usual basis for spheroidal modes: the diagonal entries are
\begin{align}
-\omega^2 + 2m\omega\Omega-m^2\Omega^2 + \omega_p^2 = (\omega_p+m\Omega-\omega)(\omega_p-m\Omega+\omega)
\label{eq:diag}
\end{align}
for the $2\ell_p+1$ $p$-modes with $m=-\ell_p...+\ell_p$, and similarly for the $g$-modes. Therefore the unperturbed solutions are $\pm\omega_{p,g}+m\Omega$, as one would expect. For the perturbations, we note that the correction $\delta{\bf A}(\omega) = {\bf R}^\dagger(0) \delta{\cal M}{\bf R}(0)$ does not depend on $\omega$; our task is ``simply'' to add this correction, re-compute the determinant ${\bf A}(\omega)$, and find the new zeroes.

We are now in a position to determine what the tidal perturbation does to the eigenfrequencies in the co-rotating frame. In order for the roots to leave the real axis, the perturbation due to the motion must first cause a pair of roots to collide and then split into a complex conjugate pair. For a general problem of the form of Eq.~(\ref{eq:ev}), we can understand this phenomenon by taking two nearby roots of the unperturbed problem, say $\omega_e$ and $\omega_d$, which correspond to zeroes in the diagonal elements $A_{ee}^{(0)}(\omega)$ and $A_{dd}^{(0)}(\omega)$ respectively. Two cases present themselves -- that $e$ and $d$ correspond to different modes (i.e. are zeroes of distinct diagonal elements), or the same mode (i.e. are zeroes of the same diagonal element, which has very nearly a repeated root). Our approach is to compute $\det{\bf A}(\omega)$ to second order in the detuning $\omega_e-\omega_d$, the separation of $\omega$ from the natural frequencies $\omega-\omega_{e,d}$, and the perturbation $\delta{\cal M}$. We then ask whether the resulting polynomial possesses roots in the upper half complex plane.

\subsection{Case of distinct modes, $e\neq d$}

In this case, we further suppose that $\omega'_e$ and $\omega'_d$ are the alternative roots associated with the same diagonal entries $A^{(0)}_{ee}$ and $A^{(0)}_{dd}$. To do this, we take $\omega-\omega_e$, $\omega-\omega_d$, and $\delta{\bf A}$ as perturbations and expand to second order. The determinant is given by the usual formula,
\begin{align}
\det{\bf A}(\omega) = \sum_\pi (-1)^\pi A_{1\pi(1)}(\omega) A_{2\pi(2)}(\omega) ... A_{N\pi(N)}(\omega),
\label{eq:det}
\end{align}
where the sum is over the $N!$ permutations $\pi$ of $\{1...N\}$ and $(-1)^\pi$ denotes whether the permutation is odd or even. Inspection shows that if at zeroth order (in $\omega-\omega_e$, $\omega-\omega_d$, and $\delta{\bf A}$) ${\bf A}(\omega)$ is diagonal and has zero entries in the $ee$ and $dd$ slots, then $\det{\bf A}(\omega)$ is second-order and the only terms at second order have $\pi(h)=h$ for $h\notin\{e,d\}$. Then $\det{\bf A}(\omega)$ is the product of the diagonal entries with $h\notin\{e,d\}$, times the determinant of the $2\times 2$ sub-block,
\begin{align}
{\bf A}(\omega) \ni \left(\begin{array}{cc}
(\omega-\omega_e)(\omega'_e-\omega) + \delta A_{ee} & \delta A_{ed} \\ \delta A_{de} & (\omega-\omega_d)(\omega'_d-\omega)+\delta A_{dd}
\end{array}\right).
\label{eq:2x2sub}
\end{align}
Moreover, at second order $\omega'_e-\omega$ in the above determinant may be replaced with $\omega'_e-\omega_e$. The latter gives a quadratic equation for $\omega$:
\begin{align}
0 = (\omega'_e-\omega_e)(\omega'_d-\omega_d)(\omega-\omega_e)(\omega-\omega_d) + (\omega'_d-\omega_d) \delta A_{ee} (\omega-\omega_d)
+ (\omega'_e-\omega_e) \delta A_{dd} (\omega-\omega_e) + \delta A_{ee}\delta A_{dd} - \delta A_{ed}^2,
\end{align}
and from the discriminant (treating the above equation as a quadratic in $\omega-\omega_e$) we get the criterion for an instability:
\begin{align}
(4\varpi_e\varpi_d\Delta - 2\varpi_d \delta A_{ee} - 2 \varpi_e \delta A_{dd})^2
-16\varpi_e\varpi_d(\delta A_{ee}\delta A_{dd} - 2\varpi_d \delta A_{ee} \Delta - \delta A_{ed}^2)
<0,
\end{align}
where we have defined $\Delta = \omega_e-\omega_d$ and $2\varpi_e=\omega_e-\omega'_e$ (so that $\varpi_e = \pm \omega_p$ for $p$-modes and $\pm\omega_g$ for $g$-modes; in general $\varpi$ denotes an inertial-frame unperturbed frequency). Algebraic simplification leads to
\begin{align}
\left(\Delta + \frac{ \delta A_{ee} }{2\varpi_e} - \frac{ \delta A_{dd}}{2\varpi_d}  \right)^2
+ \frac{\delta A_{ed}^2}{\varpi_e\varpi_d}
<0.
\end{align}
This is in fact the familiar criterion for parametric resonance. It requires first that $\varpi_e$ and $\varpi_d$ have opposite signs; if we take $\varpi_e$ to be positive, then $\omega_e = \varpi_e + m_e\Omega$, $\omega_d = \varpi_d + m_d\Omega$, and hence the resonance criterion is $|\varpi_e| + |\varpi_d| \approx (m_d-m_e)\Omega$, thus it occurs only when the unperturbed frequencies sum to a harmonic of the orbital frequency. The coupling strength must be at least $|\delta A_{ed}|>|\varpi_e\varpi_d|^{1/2}|\Delta|$, with a correction to the detuning if the applied perturbation leads to first-order corrections to the mode frequencies ($\delta A_{ee}$ and $\delta A_{dd}$).

\subsection{Case of the same mode, $e=d$}

This time we are interested in the case where two frequencies associated with the same mode are ``near'' each other and may merge -- say $\omega_e$ and $\omega'_e$. Here the frequency difference is $2\varpi_e=\omega_e-\omega'_e$ and is treated as small; we have $\omega_e = m_e\Omega + \varpi_e$ and $\omega'_e = m_e\Omega-\varpi_e$. We suppose that $\omega$ is near these frequencies, i.e. $\omega\approx m_e\Omega$, and we work to second order in $\varpi_e$, $\omega-m_e\Omega$, and $\delta{\cal M}$.

Evaluation of $\det{\bf A}(\omega)$ is more subtle than the case of $e\neq d$ because many more terms in the determinant are important. In Eq.~(\ref{eq:det}), we have two types of terms -- those with $\pi(e)=e$ and those with $\pi(e)\neq e$. Since $A_{ee}(\omega)$ is at least first order, as are all off-diagonal entries, the only surviving term with $\pi(e)=e$ is where $\pi$ is the identity permutation (i.e. corresponding to the product of all diagonal entries in ${\bf A}$). If $\pi(e) = h\neq e$, then $A_{eh}$ is first order, and so such a term can only survive if there is at most one other off-diagonal entry in the product, i.e. if $\pi(h)=e$ and $\pi(i)=i$ for all $i\notin\{e,h\}$. These terms lead to the approximation
\begin{align}
\det{\bf A}(\omega) \approx A_{ee}(\omega) \prod_{i\neq e} A_{ii}(\omega) - \sum_{h\neq e} \delta A_{eh}^2 \prod_{i\notin \{e,h\}} A_{ii}(\omega).
\end{align}
We now set this to zero and divide by $\prod_{i\neq e} A_{ii}(\omega)$ to get
\begin{align}
0 \approx A_{ee}(\omega) - \sum_{h\neq e} \frac{\delta A_{eh}^2}{A_{hh}(\omega)}.
\end{align}
Finally we see that $A_{hh}(\omega)$ can be approximated by its zeroth-order value (since it already appears multiplying a second-order perturbation), which is $(\omega-\omega_h)(\omega'_h-\omega) \approx (m_e\Omega-\omega_h)(\omega'_h- m_e\Omega)$. Also we substitute for $A_{ee}(\omega)$:
\begin{align}
0 \approx \varpi_e^2 -(\omega - m_e\Omega)^2 + \delta A_{ee} - \sum_{h\neq e} \frac{\delta A_{eh}^2}{(m_e\Omega-\omega_h)(\omega'_h- m_e\Omega)}.
\end{align}
Finally, replacing $\omega_h$ and $\omega'_h$ by $\pm\varpi_h + m_h\Omega$, we see that the roots of this equation become complex -- i.e. unstable -- when
\begin{align}
\varpi_e^2 + \delta A_{ee} - \sum_{h\neq e} \frac{\delta A_{eh}^2}{\varpi_h^2 - (m_e-m_h)^2\Omega^2} = (\omega-m_e\Omega)^2 < 0.
\label{eq:mod-stability}
\end{align}
This resembles the quasi-static instability criterion including both $A_{ee}$ (which through second order includes both three-mode and four-mode couplings of mode $e$ to the tidal field) and the square of $A_{eh}$ (three-mode coupling of $e$ and $h$ to the tidal field), modified by the rotating reference frame term, $(m_e-m_h)^2\Omega^2$. In the case of interest, where $e$ is a $g$-mode and there is a perturbation coupling to $p$-modes $h$, the instability criterion becomes
\begin{align}
\omega_g^2 + \delta A_{gg} - \frac1{\omega_p^2} \sum_{h~{\rm is}~p{\rm-mode}} \delta A_{gh}^2 \left[ 1 - \frac{(m_{g}-m_h)^2\Omega^2}{\omega_p^2} \right]^{-1} = (\omega-m_{g} \Omega)^2 < 0.
\end{align}
The last term can be expanded to leading order in $\Omega^2/\omega_p^2$ to give
\begin{align}
\omega_g^2 + \delta A_{gg} - \frac1{\omega_p^2} \sum_{h~{\rm is}~p{\rm-mode}} \delta A_{gh}^2
- \frac{\Omega^2}{\omega_p^4} \sum_{h~{\rm is}~p{\rm-mode}} (m_{g}-m_h)^2\delta A_{gh}^2
= (\omega-m_{g}\Omega)^2 < 0.
\label{eq:centrifugal}
\end{align}
We see that the instability criterion -- that the left-hand side be negative -- differs from the quasi-static case only in the introduction of a ``centrifugal correction'' of order
\begin{align}
\frac{\Omega^2 \delta A_{gp}^2}{\omega_p^4} \sim 
\left( \frac{\Omega {\cal M}_{gp}}{\omega_p^2} \right)^2 \sim 
\left( \frac{\Omega \epsilon \omega_p \omega_g J^{(1)}_{pg}}{\omega_p^2} \right)^2 \sim 
\left[ \frac{\Omega \epsilon \omega_p \omega_g (\omega_p/\omega_g)}{\omega_p^2} \right]^2\sim
\epsilon^2 \Omega^2,
\end{align}
where we have used that $\delta A_{gp}\sim {\cal M}_{gp}$, used Eq.~(\ref{eq:Mpg}) for ${\cal M}_{pg}$, and the value of the Jacobian. This correction goes in the direction of de-stabilizing the star. It has a simple interpretation in the language of the 2-dimensional oscillator of \S\ref{ss:toy}: if the eigenvectors of ${\cal M}$ rotate at some rate $\dot\theta$, then a particle moving in the shallow direction in the potential well experiences a ``centrifugal force'' correction $-\dot\theta^2$ to $\omega_-^2$. In our case where the external tidal field is rotating, the shallow direction varies by an angle ${\cal O}(\epsilon)$ over a timescale $\Omega^{-1}$, hence its rotation rate squared is $\sim\epsilon^2\Omega^2$. The interpretation as such a term is clear since in Eq.~(\ref{eq:centrifugal}), the ``rotation angle'' of mode $e$ into mode $h$ is $\delta A_{eh}/\omega_p^2$, and it oscillates at a rate $(m_e-m_h)\Omega$. The second summation then represents the sum of squared angular rates.

For the highest-order $g$-modes with frequencies $\omega_g\lesssim \epsilon\Omega$, it is possible that the centrifugal term may dominate and lead to an instability whose growth timescale would be $t_{\rm cen}\sim (\epsilon\Omega)^{-1}$.

\end{appendix}

\bibliography{References/NSRefs}

\begin{thebibliography}{}
\expandafter\ifx\csname natexlab\endcsname\relax\def\natexlab#1{#1}\fi

\bibitem[{Accadia {et~al.}(2012)Accadia, Acernese, Alshourbagy, Amico,
  Antonucci, Aoudia, Arnaud, Arnault, Arun, Astone, Avino, Babusci, Ballardin,
  Barone, Barrand, Barsotti, Barsuglia, Basti, Bauer, Beauville, Bebronne,
  Bejger, Beker, Bellachia, Belletoile, Beney, Bernardini, Bigotta, Bilhaut,
  Birindelli, Bitossi, Bizouard, Blom, Boccara, Boget, Bondu, Bonelli, Bonnand,
  Boschi, Bosi, Bouedo, Bouhou, Bozzi, Bracci, Braccini, Bradaschia, Branchesi,
  Briant, Brillet, Brisson, Brocco, Bulik, Bulten, Buskulic, Buy, Cagnoli,
  Calamai, Calloni, Campagna, Canuel, Carbognani, Carbone, Cavalier, Cavalieri,
  Cecchi, Cella, Cesarini, Chassande-Mottin, Chatterji, Chiche, Chincarini,
  Chiummo, Christensen, Clapson, Cleva, Coccia, Cohadon, Colacino, Colas,
  Colla, Colombini, Conforto, Corsi, Cortese, Cottone, Coulon, Cuoco,
  D'Antonio, Daguin, Dari, Dattilo, David, Davier, Day, Debreczeni, Carolis,
  Dehamme, Fabbro, Pozzo, del Prete, Derome, Rosa, DeSalvo, Dialinas, Fiore,
  Lieto, Emilio, Virgilio, Dietz, Doets, Dominici, Dominjon, Drago, Drezen,
  Dujardin, Dulach, Eder, Eleuteri, Enard, Evans, Fabbroni, Fafone, Fang,
  Ferrante, Fidecaro, Fiori, Flaminio, Forest, Forte, Fournier, Fournier,
  Franc, Francois, Frasca, Frasconi, Freise, Gaddi, Galimberti, Gammaitoni,
  Ganau, Garnier, Garufi, G√°sp√°r, Gemme, Genin, Gennai, Gennaro,
  Giacobone, Giazotto, Giordano, Giordano, Girard, Gouaty, Grado, Granata,
  Granata, Grave, Greverie, Groenstege, Guidi, Hamdani, Hayau, Hebri, Heidmann,
  Heitmann, Hello, Hemming, Hennes, Hermel, Heusse, Holloway, Huet, Iannarelli,
  Jaranowski, Jehanno, Journet, Karkar, Ketel, Voet, Kovalik, Kowalska,
  Kreckelbergh, Krolak, Lacotte, Lagrange, Penna, Laval, Marec, Leroy,
  Letendre, Li, Lieunard, Liguori, Lodygensky, Lopez, Lorenzini, Loriette,
  Losurdo, Loupias, Mackowski, Maiani, Majorana, Magazz√π, Maksimovic,
  Malvezzi, Man, Mancini, Mansoux, Mantovani, Marchesoni, Marion, Marin,
  Marque, Martelli, Masserot, Massonnet, Matone, Matone, Mazzoni, Menzinger,
  Michel, Milano, Minenkov, Mitra, Mohan, Montorio, Morand, Moreau, Moreau,
  Morgado, Morgia, Mosca, Moscatelli, Mours, Mugnier, Mul, Naticchioni, Neri,
  Nocera, Pacaud, Pagliaroli, Pai, Palladino, Palomba, Paoletti, Paoletti,
  Paoli, Pardi, Parguez, Parisi, Pasqualetti, Passaquieti, Passuello,
  Perciballi, Perniola, Persichetti, Petit, Pichot, Piergiovanni, Pietka,
  Pignard, Pinard, Poggiani, Popolizio, Pradier, Prato, Prodi, Punturo, Puppo,
  Qipiani, Rabaste, Rabeling, R√°cz, Raffaelli, Rapagnani, Rapisarda, Re,
  Reboux, Regimbau, Reita, Remilleux, Ricci, Ricciardi, Richard, Ripepe,
  Robinet, Rocchi, Rolland, Romano, Rosi≈Ñska, Roudier, Ruggi, Russo,
  Salconi, Sannibale, Sassolas, Sentenac, Solimeno, Sottile, Sperandio, Stanga,
  Sturani, Swinkels, Tacca, Taddei, Taffarello, Tarallo, Tissot, Toncelli,
  Tonelli, Torre, Tournefier, Travasso, Tremola, Turri, Vajente, van~den Brand,
  Broeck, van~der Putten, Vasuth, Vavoulidis, Vedovato, Verkindt, Vetrano,
  V√©ziant, Vicer√©, Vinet, Vilalte, Vitale, Vocca, Ward, Was, Yamamoto,
  Yvert, Zendri, \& Zhang}]{aVIRGO2012}
Accadia, T., Acernese, F., Alshourbagy, M., {et~al.} 2012, Journal of
  Instrumentation, 7, P03012

\bibitem[{Arun {et~al.}(2005)Arun, Iyer, Sathyaprakash, \&
  Sundararajan}]{Arun2005}
Arun, K.~G., Iyer, B.~R., Sathyaprakash, B.~S., \& Sundararajan, P.~A. 2005,
  Phys. Rev. D, 71, 084008

\bibitem[{{Bildsten} \& {Cutler}(1992)}]{Bildsten1992}
{Bildsten}, L., \& {Cutler}, C. 1992, \apj, 400, 175

\bibitem[{{Bildsten} {et~al.}(1996){Bildsten}, {Ushomirsky}, \&
  {Cutler}}]{Bildsten96}
{Bildsten}, L., {Ushomirsky}, G., \& {Cutler}, C. 1996, \apj, 460, 827

\bibitem[{Blandford \& Thorne(2011)}]{BlandfordThorne}
Blandford, R.~D., \& Thorne, K.~S. 2011, Applications of Classical Physics
  (California Institute of Technology)

\bibitem[{{Burkart} {et~al.}(2012){Burkart}, {Quataert}, {Arras}, \&
  {Weinberg}}]{Burkart12}
{Burkart}, J., {Quataert}, E., {Arras}, P., \& {Weinberg}, N.~N. 2012, ArXiv
  e-prints, arXiv:1211.1393

\bibitem[{{Chabanat} {et~al.}(1998){Chabanat}, {Bonche}, {Haensel}, {Meyer}, \&
  {Schaeffer}}]{1998NuPhA.635..231C}
{Chabanat}, E., {Bonche}, P., {Haensel}, P., {Meyer}, J., \& {Schaeffer}, R.
  1998, Nuclear Physics A, 635, 231

\bibitem[{{Cox}(1980)}]{Cox}
{Cox}, J.~P. 1980, Theory of Stellar Pulsation (Princeton University Press,
  Princeton, NJ)

\bibitem[{{Cutler} \& {Flanagan}(1994)}]{Cutler1994}
{Cutler}, C., \& {Flanagan}, {\'E}.~E. 1994, \prd, 49, 2658

\bibitem[{{Cutler} {et~al.}(1993){Cutler}, {Apostolatos}, {Bildsten}, {Finn},
  {Flanagan}, {Kennefick}, {Markovic}, {Ori}, {Poisson}, \&
  {Sussman}}]{Cutler1993}
{Cutler}, C., {Apostolatos}, T.~A., {Bildsten}, L., {et~al.} 1993, Physical
  Review Letters, 70, 2984

\bibitem[{{Deprit}(1969)}]{1969CeMec...1...12D}
{Deprit}, A. 1969, Celestial Mechanics, 1, 12

\bibitem[{Flanagan \& Hinderer(2008)}]{FlanaganHinderer2008}
Flanagan, E.~E., \& Hinderer, T. 2008, Phys. Rev. D, 77, 021502

\bibitem[{{Fuller} \& {Lai}(2011)}]{Fuller11}
{Fuller}, J., \& {Lai}, D. 2011, \mnras, 412, 1331

\bibitem[{{Fuller} \& {Lai}(2012)}]{Fuller12}
---. 2012, \mnras, 421, 426

\bibitem[{{Fuller} \& {Lai}(2013)}]{Fuller13}
---. 2013, \mnras, 430, 274

\bibitem[{{Harry} \& the LIGO Scientific~Collaboration(2010)}]{aLIGO2010}
{Harry}, G., \& the LIGO Scientific~Collaboration. 2010, Classical and Quantum
  Gravity, 27, 084006

\bibitem[{Hinderer {et~al.}(2010)Hinderer, Lackey, Lang, \&
  Read}]{HindererLackey2010}
Hinderer, T., Lackey, B.~D., Lang, R.~N., \& Read, J.~S. 2010, Phys. Rev. D,
  81, 123016

\bibitem[{Hotokezaka {et~al.}(2013)Hotokezaka, Kyutoku, \&
  Shibata}]{Hotokezaka2013}
Hotokezaka, K., Kyutoku, K., \& Shibata, M. 2013, Phys. Rev. D, 87, 044001

\bibitem[{Kiuchi {et~al.}(2010)Kiuchi, Sekiguchi, Shibata, \&
  Taniguchi}]{Kiuchi2010}
Kiuchi, K., Sekiguchi, Y., Shibata, M., \& Taniguchi, K. 2010, Phys. Rev.
  Lett., 104, 141101

\bibitem[{{Kochanek}(1992)}]{Kochanek1992}
{Kochanek}, C.~S. 1992, \apj, 398, 234

\bibitem[{{Lai}(1994)}]{Lai1994}
{Lai}, D. 1994, \mnras, 270, 611

\bibitem[{{Lai} {et~al.}(1994{\natexlab{a}}){Lai}, {Rasio}, \&
  {Shapiro}}]{LaiRasio1994b}
{Lai}, D., {Rasio}, F.~A., \& {Shapiro}, S.~L. 1994{\natexlab{a}}, \apj, 423,
  344

\bibitem[{{Lai} {et~al.}(1994{\natexlab{b}}){Lai}, {Rasio}, \&
  {Shapiro}}]{LaiRasio1994a}
---. 1994{\natexlab{b}}, \apj, 420, 811

\bibitem[{{Lynden-Bell} \& {Ostriker}(1967)}]{LyndenBell1967a}
{Lynden-Bell}, D., \& {Ostriker}, J.~P. 1967, \mnras, 136, 293

\bibitem[{{McDermott} {et~al.}(1988){McDermott}, {van Horn}, \&
  {Hansen}}]{1988ApJ...325..725M}
{McDermott}, P.~N., {van Horn}, H.~M., \& {Hansen}, C.~J. 1988, \apj, 325, 725

\bibitem[{{McDermott} {et~al.}(1983){McDermott}, {van Horn}, \&
  {Scholl}}]{1983ApJ...268..837M}
{McDermott}, P.~N., {van Horn}, H.~M., \& {Scholl}, J.~F. 1983, \apj, 268, 837

\bibitem[{Oechslin \& Janka(2007)}]{Oechslin2007}
Oechslin, R., \& Janka, H.-T. 2007, Phys. Rev. Lett., 99, 121102

\bibitem[{{Read} {et~al.}(2013){Read}, {Baiotti}, {Creighton}, {Friedman},
  {Giacomazzo}, {Kyutoku}, {Markakis}, {Rezzolla}, {Shibata}, \&
  {Taniguchi}}]{Read2013}
{Read}, J.~S., {Baiotti}, L., {Creighton}, J.~D.~E., {et~al.} 2013, ArXiv
  e-prints, arXiv:1306.4065

\bibitem[{{Reisenegger} \& {Goldreich}(1992)}]{Reisenegger1992}
{Reisenegger}, A., \& {Goldreich}, P. 1992, \apj, 395, 240

\bibitem[{Schenk {et~al.}(2002)Schenk, Arras, Flanagan, Teukolsky, \&
  Wasserman}]{Schenk2002}
Schenk, A.~K., Arras, P., Flanagan, E.~E., Teukolsky, S.~A., \& Wasserman, I.
  2002, Phys.Rev., D65, 024001

\bibitem[{Somiya(2012)}]{KAGRA2012}
Somiya, K. 2012, Classical and Quantum Gravity, 29, 124007

\bibitem[{{Steiner} \& {Watts}(2009)}]{2009PhRvL.103r1101S}
{Steiner}, A.~W., \& {Watts}, A.~L. 2009, Physical Review Letters, 103, 181101

\bibitem[{{the LIGO Scientific Collaboration} {et~al.}(2013){the LIGO
  Scientific Collaboration}, {the Virgo Collaboration}, {Aasi}, {Abadie},
  {Abbott}, {Abbott}, {Abbott}, {Abernathy}, {Accadia}, {Acernese}, \&
  et~al.}]{LIGOParam2013}
{the LIGO Scientific Collaboration}, {the Virgo Collaboration}, {Aasi}, J.,
  {et~al.} 2013, ArXiv e-prints, arXiv:1304.1775

\bibitem[{{Thorne}(1969{\natexlab{a}})}]{1969ApJ...158....1T}
{Thorne}, K.~S. 1969{\natexlab{a}}, \apj, 158, 1

\bibitem[{{Thorne}(1969{\natexlab{b}})}]{1969ApJ...158..997T}
---. 1969{\natexlab{b}}, \apj, 158, 997

\bibitem[{{Thorne} \& {Campolattaro}(1967)}]{1967ApJ...149..591T}
{Thorne}, K.~S., \& {Campolattaro}, A. 1967, \apj, 149, 591

\bibitem[{{Tsang} {et~al.}(2012){Tsang}, {Read}, {Hinderer}, {Piro}, \&
  {Bondarescu}}]{2012PhRvL.108a1102T}
{Tsang}, D., {Read}, J.~S., {Hinderer}, T., {Piro}, A.~L., \& {Bondarescu}, R.
  2012, Physical Review Letters, 108, 011102

\bibitem[{{Unno} {et~al.}(1989){Unno}, {Osaki}, {Ando}, {Saio}, \&
  {Shibahashi}}]{Unno1989}
{Unno}, W., {Osaki}, Y., {Ando}, H., {Saio}, H., \& {Shibahashi}, H. 1989,
  Nonradial Oscillations of Stars (University of Tokyo Press, Tokyo)

\bibitem[{{Van Hoolst}(1994)}]{VanHoolst1994}
{Van Hoolst}, T. 1994, \aap, 286, 879

\bibitem[{{Weinberg} {et~al.}(2013){Weinberg}, {Arras}, \&
  {Burkart}}]{Weinberg2013}
{Weinberg}, N.~N., {Arras}, P., \& {Burkart}, J. 2013, \apj, 769, 121

\bibitem[{Weinberg {et~al.}(2012)Weinberg, Arras, Quataert, \&
  Burkart}]{Weinberg2012}
Weinberg, N.~N., Arras, P., Quataert, E., \& Burkart, J. 2012, Astrophys.J.,
  751, 136

\bibitem[{Wu \& Goldreich(2001)}]{Wu2001}
Wu, Y., \& Goldreich, P. 2001, Astrophys.J., 546, 469

\end{thebibliography}

\end{document}